\documentclass{aa}  
\usepackage{comment}
\usepackage{siunitx}
\usepackage{subcaption}
\usepackage{nicefrac}
\usepackage{graphicx}
\usepackage{color}
\usepackage{hyperref}
\hypersetup{colorlinks,linkcolor={blue},citecolor={blue}}  
\usepackage[varg]{txfonts}
\begin{document} 

    \title{Emerging population of gap-opening planets around type-A stars}
    \subtitle{Long-term evolution of the forming planets around HD 163296}
   %\title{The gas giant population emerging from embedded planets}

   \author{J. M\"uller-Horn \inst{1,2} \and
          G. Pichierri \inst{1} \and
          B. Bitsch \inst{1}}

   \institute{Max-Planck-Institut f\"ur Astronomie, K\"onigstuhl 17, 69117 Heidelberg, Germany
         \and
             Institut für Astrophysik, Georg-August-Universit\"at, Friedrich-Hund-Platz 1, 37077 G\"ottingen, Germany
              \thanks{email: j.muellerhorn@stud.uni-goettingen.de}
             }
   \date{Received 14 February 2022 / Accepted 21 April 2022}

% \abstract{}{}{}{}{} 
% 5 {} token are mandatory
 
  \abstract
  % context heading (optional)
  % {} leave it empty if necessary  
   {The presence of forming planets embedded in their protoplanetary disks has been inferred from the detection of multiring structures in such disks. Most of these suspected planets are undetectable by direct imaging observations at current measurement sensitivities. Inward migration and accretion might make these putative planets accessible to the Doppler method, but the actual extent of growth and orbital evolution remains unconstrained.}
  % aims heading (mandatory)
   {Under the premise that the gaps in the disk around HD 163296 originate from new-born planets, we investigate if and under which circumstances the gap-opening planets could represent progenitors of the exoplanet population detected around A-type stars. In particular, we study the dependence of final planetary masses and orbital parameters on the viscosity of the disk.}
  % methods heading (mandatory)
   {The evolution of the embedded planets was simulated throughout the disk lifetime and up to 100\,Myr after the dispersal of the disk, taking the evolving disk structure and a likely range of disk lifetimes into account. We modeled the temperature and density structure of the disk based on observational results following the widely used $\alpha$ disk prescription and we varied planetary and disk parameters within the available observational constraints.}
  % results heading (mandatory)
   {We find that the final configuration of the planets is largely determined by the $\alpha$ viscosity parameter of the disk and less dependent on the choice for the disk lifetime and the initial planetary parameters. If we assume that planets such as those in HD 163296 evolve to form the observed exoplanet population of A-type stars, a $\alpha$ parameter on the order of $\num{3.16e-4} \lesssim \alpha \lesssim 10^{-3}$ is required for the disks to induce sufficiently high migration rates. Depending on whether or not future direct imaging surveys will uncover a larger number of planets with $m_\mathrm{pl} \lesssim 3\,M_\mathrm{Jup}$ and $a_\mathrm{pl} \gtrsim 10\,\mathrm{AU}$, we expect the $\alpha$ parameter to be at the lower or upper end of this range, always under the assumption that such disks indeed harbor wide orbit planets.}
   {}


   \maketitle
%
%-------------------------------------------------------------------

\section{Introduction}
Detailed observations by the Atacama Large Millimeter/submillimeter Array (ALMA) have revealed a high substructure level in the protoplanetary disks around young stars \citep{ALMA}. These substructures frequently present themselves in the form of nearly axisymmetric gaps and rings in the dust continuum emission \citep{Walsh, Andrews, Fedele}. A common interpretation of these structures is that they originate from the interaction between the disk and massive planets embedded therein \citep{Long, DSHARP-Huang, Pinilla}. A sufficiently massive planet increases the gas-velocity exterior to its orbit to super-Keplerian speeds and in doing so generates a pressure bump in the disk. The pressure bump stops the inward drift of pebbles and the particles accumulate outside the planet's orbit \citep{Paardekooper2006}, thus providing an explanation for the observed ring- and gap-like structures. These models were shown to be capable of reproducing the observed disk emission using hydrodynamical and radiative transfer simulations \citep{Dipierro, Clarke, Nazari}. 

If the observed rings are in fact induced by gap-opening planets, then we might expect these embedded planets to eventually form the massive planet population detected by the extensive exoplanet surveys of the last years \citep{Wittenmyer2020, Fulton}. However, radial distances of known exoplanets lie in the range 0.1-10\,AU as opposed to typical radii of the rings in disks that are of on the order of 10-100\,AU \citep{DSHARP-Huang}. With few exceptions from direct imaging, this region is not yet accessible with current exoplanet detection techniques. Hence, the question arises of whether the yet unseen planets are in fact the progenitors of the discovered exoplanet population. This would require sufficient orbital evolution and inward migration to bring them closer to the star and into the observable regime. From the perspective of planet formation theory, it is therefore interesting to study how these presumed embedded planets would evolve and to compare the emerging population with the one of known exoplanets. 

The question of how the suspected embedded planets will evolve has been addressed by \citet{Lodato} who studied a large sample of 48 gaps in 32 protoplanetary disks and approximated the masses of the suspected planets from the gap morphology. After evolving the planets for 3-5\,Myr, they found a good agreement between the final configuration of the simulated planets and the distribution of cold Jupiters. The authors note, however, that the evolutionary model they used is very simplified and that several aspects could be improved. Caveats include the fact that the disk structure remained fixed throughout the integration and that the same generic lifetimes, disk profiles and viscosities were assumed for all disks in the sample. Furthermore, only the one-planet case has been investigated neglecting possible planet-planet interaction in multiple planet systems. The same is true for a similar study carried out by \citet{Ndugu}, who pointed out that the disk viscosity is indeed key in shaping the final planet population and that the low viscosities needed to explain the core growth at these large distances are not consistent with inward migration from several 10 AUs to sub 10 AU orbits. 

In this paper, we choose an opposite approach where instead of simulating a large sample of disks we investigate the evolution of planets in only one disk, but in turn pay closer attention to the properties of the disk structure as well as considering multiplanet systems. Furthermore, we take care to motivate our initial conditions exclusively from observational constraints. 

Our focus lies on the planet candidates of Herbig Ae star HD 163296 whose existence has been surmised both from hydrodynamical simulations and gas kinematics in the disk. With a radius of more than 500 AU inferred from CO line emission \citep{Isella+2017}, the protoplanetary disk around HD 163296 is one of the largest known disks with resolved substructure. It is located at a distance of $101\,$pc \citep{Gaia} and has an estimated system age of approximately $5\,$Myr \citep{Montesinos}. Due to the likely presence of forming planets the disk has been the subject of extensive research in recent years. Observations revealed a high degree of substructure in the spatial distribution of millimeter-sized particles and cold molecular gas. These substructures include the presence of 4 gaps at radii of $\sim 10$, 48, 86, and 145\, AU and a crescent-shaped asymmetry at 55\, AU which have been inferred from millimeter continuum and CO molecular line emission \citep{Isella+2016, DSHARP-Huang, Isella+2018}. Hydrodynamical simulations point toward a likely planetary origin of these features \citep{Zhang2018, Liu, Rodenkirch} and recent detections of meridional flows and deviations from Keplerian rotation \citep{Teague2018,Pinte, Teague19} provide further evidence of active planet-disk interaction. 

Assuming the gaps in HD 163296 are indeed of planetary origin, the system represents a direct example of the initial stages of planet formation. For the purpose of this paper, we adopt this hypothesis to investigate the future evolution of the planetary system. We would like to point out, however, that this interpretation is not undisputed and several other mechanisms have been suggested to account for the presence of these features, among other things particle trapping at the outer edge of dead zones \citep{Ruge}, pebble growth around condensation fronts \citep{Zhang+2015}, self-induced dust pile-ups \citep{Gonzalez}, as well as zonal flows \citep{Flock2015}. 

The paper is organised as follows; in Section \ref{sec:methods} we describe our disk model, the applied accretion and migration schemes and the setup of simulations. In Section \ref{sec:results} we show the results for two exemplary individual systems and present the ensemble statistics of the simulated planet population. We summarize our findings and discuss possible implications in Sections \ref{sec:discussion} and \ref{sec:conclusion}. 

%--------------------------------------------------------------------
\section{Methods}
\label{sec:methods}
Our simulations were performed using the planet formation code \texttt{FLINTSTONE}. The code models the evolution of planets in a gas-dominated protoplanetary disk thereby taking planetary migration, accretion, the gas damping of eccentricities and inclinations, as well as the evolution of the disk structure into account. The orbital integration was carried out by the \texttt{MERCURY} hybrid symplectic integrator which resolves close encounters \citep{mercury}. In our simulations, we treated collisions between planets as inelastic mergers. A similar version of the code was used by \citet{Bitsch19} and \citet{Izidoro21} so we refer to them for a more detailed description and will limit ourselves here to a summary of the main concepts and to highlight any differences. 

\subsection{Disk model}
The structure of the protoplanetary disk determines the growth and evolution of the planets it contains and is characterized by physical properties such as its geometry, temperature, and density distribution. In contrast to \citet{Ndugu} who used disk profiles derived from hydrodynamical simulations and \citet{Lodato} who used a generic fixed disk profile for all disks in their sample, here we focused only on one disk, namely the one around HD 163296, and aimed to model its structure based on the available observational data. For this purpose, we followed \citet{Kama} who constrained the disk gas mass of HD 163296 using hydrogen deuteride rotational line emission. Their estimate of the total gas mass amounts to $M_{\mathrm{gas}} = 0.067\,M_{\odot}$. If not indicated otherwise, we used the disk parameters obtained by \citet{Kama} which are summarized in Table \ref{tab:discparams}.

\begin{table}[!t]
\caption{Adopted model parameters for the disk of HD 163296 \citet{Kama}.}            
\label{tab:discparams}      
\centering                          % used for centering table
\begin{tabular}{c c}       
\hline\hline                 
Parameter & Value  \\    % table heading 
\hline                        
   $\gamma$ & 0.9 \\
   $\Psi$ & 0.05  \\
   $h_{\mathrm{c}}$ & 0.075 \\
   $r_{\mathrm{c}}$ [AU] & 125  \\ 
   $M_{\mathrm{gas}}$ [$M_{\odot}$] & 0.067 \\
   $L_* [L_{\odot}]$ & 37.7 \\
%   $r_{in}$ [AU] & 0.7  \\ 
\hline                                   %inserts single line
\end{tabular}
\end{table}

Following the viscous accretion formalism \citep{Lynden,Hartmann}, the initial distribution of gas in the disk is described by the vertically integrated surface density profile $\Sigma_{\mathrm{gas},0}$ as
\begin{equation}
    \Sigma_{\mathrm{gas},0}(r) = \Sigma_{\mathrm{c}} \left(\frac{r}{r_{\mathrm{c}}}\right)^{-\gamma} \exp{\left[\left(-\frac{r}{r_{\mathrm{c}}}\right)^{2-\gamma}\right]}
\end{equation}
where $\Sigma_{\mathrm{c}} = (2-\gamma) \frac{M_{\mathrm{gas}}}{2 \pi r_{\mathrm{c}}^2}$ is the surface density at the characteristic radius $r_{\mathrm{c}} = 125\,$AU and $\gamma$ is the power-law index. At the characteristic radius $r_c = 125~ \mathrm{AU}$ this corresponds to an initial surface density value of $\Sigma_{\mathrm{gas},0}(r = r_c) = 1.29~ \frac{\mathrm{g}}{\mathrm{cm}^2}$. Assuming a locally isothermal state the temperature profile can be expressed as
\begin{equation}
    T(r) = \frac{\mu m_{\mathrm{p}}}{k_{\mathrm{B}}} \, \frac{G M_*}{r} \,h(r)^2 \label{eq:temperature}
\end{equation}
with the proton mass $m_{\mathrm{p}}$, the gravitational constant $G$, the mass of the central star $M_* = 1.9\,M_{\odot}$ \citep{Rodenkirch}, and a mean molecular weight $\mu$ which we set to $2.3$. In Eq. \eqref{eq:temperature} we introduced the aspect ratio $h(r) = \frac{H}{r}$ with $H$ being the disk scale height. For the aspect ratio we used a power-law parameterization 
\begin{equation}
    h(r) = \frac{c_{\mathrm{s}}}{v_K} = h_{\mathrm{c}} \left(\frac{r}{r_{\mathrm{c}}}\right)^{\Psi}.
\end{equation}
In this equation $c_{\mathrm{s}}$ is the sound speed, $v_K$ the Keplerian orbital velocity, $h_{\mathrm{c}}$ the opening angle at the characteristic radius $r_{\mathrm{c}}$ and $\Psi$ is the flaring index of the disk. With $\Psi = 0.05$, as found by \citet{Kama}, the disk is slightly less flared than expected from theory; \citet{Chiang} derived a flaring index of $\Psi_{\mathrm{theo}} = 2/7$ for flared disks in radiative equilibrium. A lower flaring index results in a larger torque exerted onto the planets which in turn leads to a shorter migration timescale (see Sect. \ref{sec:migration}). We would like to mention here that a recently published study by \citet{Calahan} reinvestigated the thermal structure of HD 163296's disk and finds updated parameters for its profile. In our notation their findings correspond to parameter values of $h_{\mathrm{c}}=0.087$, $r_{\mathrm{c}}=165$\,AU, $M_{\mathrm{gas}} = 0.14\,M_{\odot}$ and a flaring index $\Psi=0.08$. These results are in good agreement with the values derived by \citet{Kama} with the exception of a higher total gas mass, $M_{\mathrm{gas}}$. We do, however, not expect this difference to influence our results significantly, see Sect. \ref{sec:discussion} for further details.

In all simulations, the disk extended from $r_{\mathrm{in}} = 0.7$\,AU to $r_{\mathrm{out}} = 300$\,AU where the inner edge of the disk was chosen in accordance with \citet{Flock} to account for an inner cavity. As was done in \citet{Izidoro21}, we parameterized the drop in surface density in the inner region of the disk by multiplying the gas density with and additional factor $\mathfrak{R}$
\begin{equation}
    \mathfrak{R} = \tanh{\left(\frac{r-r_{\mathrm{in}}}{C r_{\mathrm{in}}}\right)}
\end{equation}
with $C = 0.05$ to allow for a smooth transition. The outer boundary represents the applied ejection distance and was chosen for computational reasons. 

\begin{figure}[!t]
   \centering
   \includegraphics[width=\hsize]{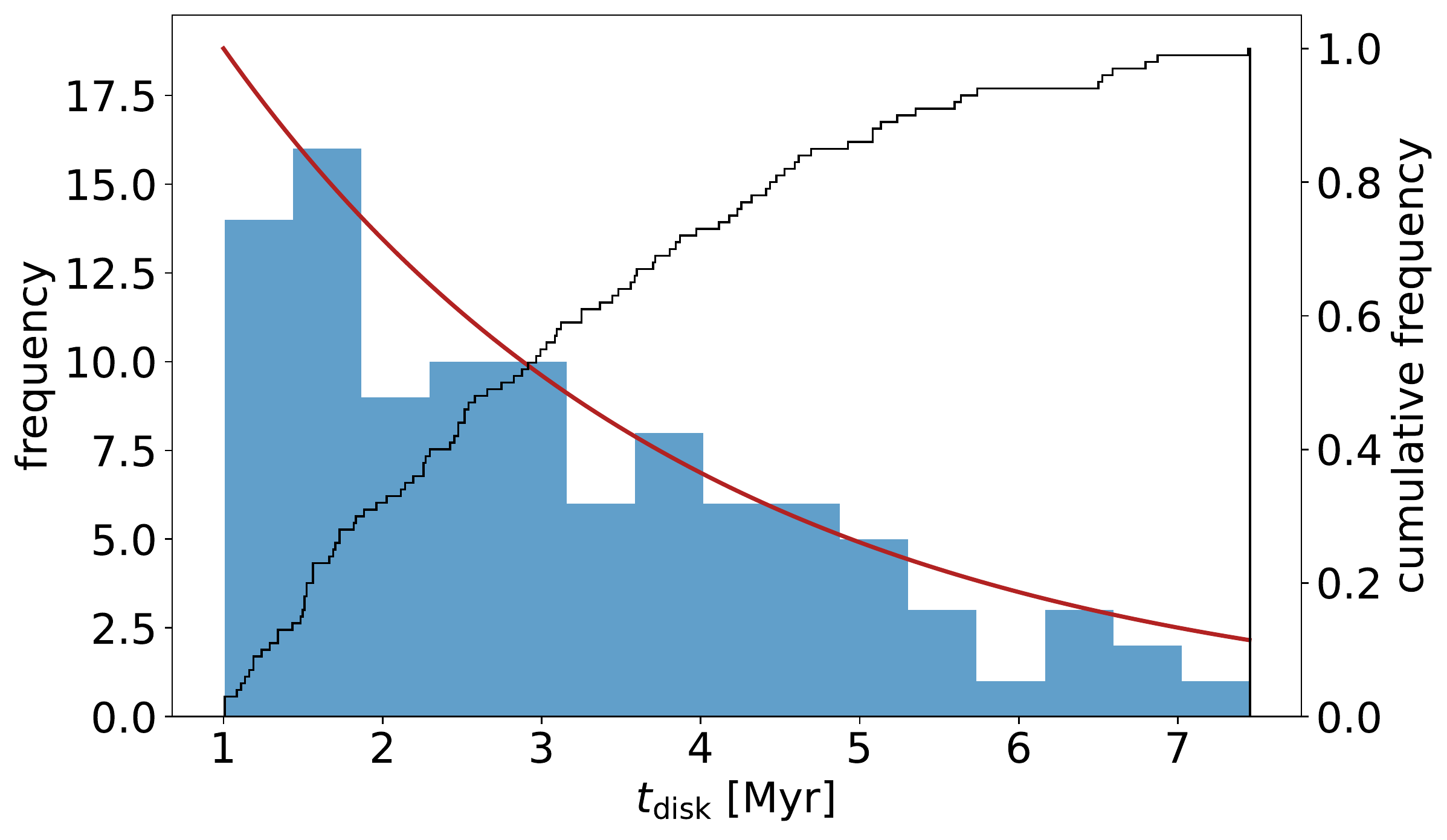}
      \caption{Distribution of gas disk lifetimes $t_{\mathrm{disk}}$ used in the simulations. The 100 samples (blue histogram) drawn randomly from the underlying exponential distribution (red line) are displayed as well as the cumulative frequency (black outline). The distribution is limited to lifetimes between $1 \leq t_{\mathrm{disk}} \leq 7.5$\,Myr additional to the current age of the system of approximately $5$\,Myr.}
         \label{fig:gaslifetime}
 \end{figure}

Following \citet{Shakura}, we used an $\alpha$ disk model to parameterize the turbulent viscosity $\nu$ of the disk. In this widely used prescription, the parameter $\alpha$ is linked to the effective viscosity of the disk via $\nu = \alpha  c_{\mathrm{s}} H$. There exist observational constraints on $\alpha$ in the literature but they vary widely in order of magnitude. The distribution of $\alpha$ values inferred from disk observations by ALMA spans a range from $10^{-4}$ to $4 \times 10^{-2}$ \citep{Rafikov, Dullemond}. For the particular case of the HD 163296 disk, \citet{Flaherty} place an upper limit on the turbulent viscosity corresponding to $\alpha \leq 3 \times 10^{-3}$. Recently, \citet{Rodenkirch} derived a refined upper bound of  $\alpha \leq 2 \times 10^{-3}$ which they find is needed to account for an asymmetric feature in the inner disk. Due to the prevailing uncertainty regarding the precise value, we left $\alpha$ as a free parameter of the study, varying it between $10^{-4}$ and $10^{-3}$.

The disk structure is, however, not static but evolves on a viscous timescale mainly due to accretion onto the central star. In our simulations, the surface density decayed exponentially on a $\tau_{\mathrm{diss,visc}} = 3$\,Myr timescale to mimic gas dissipation \citep{Mamajek}. In order to save computation time, the disk structure was recalculated only every 500\,yr rather than for every time step of the numerical integrator. Since the disk evolution takes place on Myr timescales this measure does not affect the validity of our results. The total gas lifetime $t_{\mathrm{disk}}$ was treated as a free parameter in the simulations and samples were randomly drawn from an exponential distribution as suggested by \citet{Mamajek}. We set the lower limit of the gas lifetime to $1$\,Myr in addition to the current age of the system of $t_0 = 5$\,Myr and cut off the high-end tail of the distribution above $t_0 + 7.5$\,Myr to limit the required computational resources. This setup corresponds to total disk lifetimes ranging from $6$ to $12.5$\,Myr in accordance with the findings of \citet{Michel}. Since only on the order of 10\% of the samples lie above $7.5$\,Myr we do not expect this limit to have a significant impact on the results. The distribution of gas lifetimes in our simulations is displayed in Fig. \ref{fig:gaslifetime}. Longer disk lifetimes imply more time for the planets to grow and migrate resulting in a more massive and closer-in final population.

During the final stages of the disk lifetime the density becomes low enough for photoevaporation to set in and effectively clear the remainders of the disk on a shorter timescale \citep{Alexander, Owen}. In our simulations, this dissipation process was modeled by decreasing the surface density exponentially with a timescale of $\tau_{\mathrm{diss,ph.ev.}} = 35$\,Kyr over a period of 300\,Kyr from $t_{\mathrm{diss, ph.ev.}} = t_{\mathrm{disk}} - 300\,$Kyr until the end of gas disk life time, as in other N-body simulations \citep{Bitsch19, Izidoro21}. The time-dependent decrease in surface density due to dissipation in our model was, hence, calculated as
\begin{align}
\begin{split}
    \Sigma_{\mathrm{gas}}(r,t) = \Sigma_{\mathrm{gas},0}(r) &\exp{\left[-\left(\frac{t}{\tau_{\mathrm{diss,visc}}}\right)\right]} \cdot \\
                                     &\left\{%
                                    \begin{array}{ll}
                                            1, & \hbox{$t < t_{\mathrm{diss,ph.ev.}}$} \\
                                            \exp{\left[-\left(\frac{t-t_{\mathrm{diss,ph.ev.}}}{\tau_{\mathrm{diss,ph.ev.}}}\right)\right]}, & \hbox{$t>t_{\mathrm{diss,ph.ev.}}$} \\
                                    \end{array}%
                                    \right.
\end{split}
\end{align}
where the time $t$ is set to zero at the beginning of the simulation.

\subsection{Gas accretion}
The growth and migration of planets occurs for the duration of the disk lifetime. Planetary cores grow predominantly through the accretion of pebbles which experience an inward drift due to the aerodynamic drag \citep{Brauer}. For the more massive planets gas accretion is the dominant growth mechanism. The transition occurs when the growing planet carves a partial gap in the density of the protoplanetary disk. The radial pressure gradient exterior to the orbit of the planet gets reversed and the inward drift of pebbles is halted \citep{Morbidelli, Lambrechts14, Bitsch18, Ataiee}. Following \citet{Bitsch18}, we express the pebble isolation mass as
\begin{equation}
    M_{\mathrm{iso}} = 25 f_{\mathrm{fit}} M_\mathrm{Earth} \label{eq:Miso}
\end{equation} where
\begin{equation*}
    f_{\mathrm{fit}} = \left(\frac{h}{0.05}\right)^3 \left(0.34\left(\frac{\log 0.001}{\log \alpha }\right)^4 + 0.66 \right) \left(1-\frac{\frac{\partial \ln P}{\partial \ln r}+2.5}{6}\right)\,.
\end{equation*}
It depends on the local properties of the disk, namely the aspect ratio $h$, the pressure gradient $\frac{\partial \ln P}{\partial \ln r}$, and the $\alpha$ viscosity parameter.

The observed ringlike gaps in the protoplanetary disk of HD 163296 indicate that the inferred planets have already reached pebble isolation mass. Therefore, we assumed that their growth is driven by gas accretion. For the process of rapid gas accretion we followed \citet{Machida} who performed three-dimensional hydrodynamical simulations to investigate the mass accretion rate of Jovian planets. Their findings indicate that the effective accretion rate is given by $\Dot{M}_{\mathrm{gas}} = \min \{\Dot{M}_{\mathrm{gas, low}}, \Dot{M}_{\mathrm{gas, high}}\}$ where 
\begin{equation}
    \Dot{M}_{\mathrm{gas, low}} = 0.83 \Omega_K \Sigma_{\mathrm{gas}} H^2 \left(\frac{r_H}{H}\right)^{\frac{9}{2}} \label{eq:acc_rate1}
\end{equation} and
\begin{equation}
    \Dot{M}_{\mathrm{gas, high}} = 0.14 \Omega_K \Sigma_{\mathrm{gas}} H^2\,. \label{eq:acc_rate2}
\end{equation}
Here $r_H$ denotes the Hill radius of the planet and $\Omega_K$ is the Keplerian frequency. However, we do note that there is great uncertainty in the literature about gas accretion rates and suggested values range over several orders of magnitude \citep{Ayliffe, Machida, Angelo, Schulik}.  Ultimately, accretion is limited by what the disk can provide. \cite{Lubow} found that the mass flow rate across the gap opened by a planet lies typically between 10 and 25\% of the mass accretion rate outside the orbit of the planet. As was done by \citet{Bitsch19}, we chose to restrict the maximum accretion rate for gap-opening planets to an intermediate value of 80\% of the disk's accretion rate on the star, given by 
\begin{equation}
    \Dot{M} = 3 \pi \alpha  h^2 r^2 \Omega_K \Sigma_{\mathrm{gas}}\,. \label{eq:acc_rate3}
\end{equation}
For a large fraction of planets in our simulations this limit is actually lower than the rates of Eqs. \eqref{eq:acc_rate1} and \eqref{eq:acc_rate2}. 

For very massive, gap-opening planets the reduction in surface density at the planet's position may lead to a decrease of the accretion rate onto the planet \citep{Tanigawa2016} which was not taken into account in our simulations. We consider this choice to be justified since for planets less massive than $10\,M_{\mathrm{Jup}}$, which make up the vast majority of simulated planets, accretion is not heavily impacted by the formation of gaps \citep{Tanigawa2016} and gas can traverse the gap even for massive planets \citep{Lubow}.
Hydrodynamic simulations indicated, moreover, that massive planets can excite radial motion in the surrounding gas leading to an increase in eccentricity for the outer gap edge \citep{Kley2006, Li2021, Tanaka2022}. The onset of gap eccentricity can significantly enhance the feeding zone of the planet and in turn cause larger mass accretion rates. \citet{Kley2006} suggest that this effect becomes considerable for planet to star mass ratios of $q>0.005$, which corresponds to planet masses $m_{\mathrm{pl}} \gtrapprox 10\,M_{\mathrm{Jup}}$ in the case of HD 163296. 
Such massive planets are formed only in rare cases in our simulations, and when they are, it is toward the very end of the disk's lifetime or through collisions. For this reason we have not considered the effect of gap eccentricity on the mass accretion rates in our simulations but we would like to point out the uncertainty of the mass accretion rates for the most massive planets.

In cases where planets cross the inner edge of the disk at $0.7$\,AU and the surface density vanishes, accretion rates were set to zero and prevent further growth. Even though in principle gas accretion could continue in the inner region because the stellar accretion rate remains constant, it is unclear whether gas accretion onto planets would proceed normally \citep{Cimerman2017,Lega2019,Lambrechts19}. In light of this uncertainty and the fact that accretion at the inner edge does not play a significant role for the majority of our simulations, we set accretion to zero in this region for definiteness. 

\subsection{Orbital migration}
\label{sec:migration}
Growing planets interact with the protoplanetary disk, thereby exchanging energy and angular momentum. The torque induced on planets by the disk leads to planetary migration and the timescale thereof determines the final configurations of the planetary system. Lower-mass embedded planets are subject to migration in the type-I regime for which we followed the prescription by \citet{Paardekooper} who determined a torque formula for nonisothermal type-I migration which has been tested against 3D hydrodynamic simulations \citep{Lega2015, BitschKley}:
\begin{equation}
    \Gamma_{\mathrm{tot}} = \Delta_{\mathrm{C}} \, \Gamma_{\mathrm{C}} + \Delta_{\mathrm{L}} \, \Gamma_{\mathrm{L}}.
\end{equation}
The total torque is computed as the sum of the Lindblad torque $\Gamma_{\mathrm{L}}$ which originates from density waves launched at Lindblad resonances and the entropy-related corotation torque $\Gamma_{\mathrm{C}}$. The rescaling factors $\Delta_{\mathrm{L}}$ and $\Delta_{\mathrm{C}}$ were introduced by \citet{Izidoro17} to account for torque reduction due to the orbital eccentricity and inclination. The torque depends strongly on the local properties and gradients of the disk and scales with the square of the planetary mass. For the specific equations used for implementation we refer to \citet{Izidoro21} (see their Appendix A for the detailed equations). The timescale of type-I migration is given by $\tau_{\mathrm{mig, \MakeUppercase{\romannumeral 1}}} = - \frac{L}{\Gamma_{\mathrm{tot}}}$ with $L$ being the angular momentum of the planet.

More massive, gap-opening planets enter the slower type-II migration regime. Classically, a planet carves a gap in the protoplanetary disk if \citep{Crida} 
\begin{equation}
\label{eq:crida2006}
    \mathcal{P} = \frac{3}{4} \frac{H}{r_H} + \frac{50}{q \mathcal{R}} \leq 1 ,
\end{equation}
where $q$ denotes the star to planet mass ratio and $\mathcal{R} = r^2 \Omega_K / \nu$ is the Reynolds number. Recently, a new approximation for gap opening, especially for the transition between the two regimes, has been put forward by \citet{Kanagawa}. They relate the timescale of type-II migration $\tau_{\mathrm{mig, II}}$ to type-I migration via
\begin{equation}
    \tau_{\mathrm{mig, II}} = \frac{\Sigma_{\mathrm{un.p}}}{\Sigma_{\mathrm{min}}} \tau_{\mathrm{mig, I}} \, . \label{eq:migration_timescale}
\end{equation}
The scaling factor in this equation is given by the ratio between the unperturbed surface $\Sigma_{\mathrm{un.p}}$ and the surface density at the bottom of the gap $\Sigma_{\mathrm{min}}$ and can be expressed as \citep{Duffell, Kanagawa15}
\begin{equation}
    \frac{\Sigma_{\mathrm{un.p}}}{\Sigma_{\mathrm{min}}} = 1 + 0.04 K,~ K = \left(\frac{m_\mathrm{pl}}{M_{\odot}}\right)^2  \left(\frac{H}{r}\right)^{-5} \alpha ^{-1}, \label{eq:density_ratio}
\end{equation}
where $m_\mathrm{pl}$ denotes the planet mass. We followed here the approach by \citet{Kanagawa}. Equations \eqref{eq:migration_timescale} and \eqref{eq:density_ratio} imply that a higher viscosity of the disk causes the transition between type-I and type-II migration to occur later. As a result, the distance covered by planets through migration is larger.

Finally, the tidal damping of inclinations and eccentricities was implemented using the prescription by \citet{Cresswell06} and \citet{Cresswell08}. Specifically, in the type-I regime the eccentricity and inclination damping timescales $\tau_e$ and $\tau_i$ are given by

\begin{equation}
    \begin{split}
    \tau_e &= \frac{\tau_\mathrm{wave}}{\num{0.780}}\left[1-0.14\left(\frac{e}{H/r}\right)^2+0.06\left(\frac{e}{H/r}\right)^3\right.\\
    &\left.+0.18\left(\frac{e}{H/r}\right)\left(\frac{i}{H/r}\right)^2\right]\text{,}
    \end{split}
\end{equation}
\begin{equation}
    \begin{split}
    \tau_i &= \frac{\tau_\mathrm{wave}}{\num{0.544}}\left[1-0.30\left(\frac{i}{H/r}\right)^2+0.24\left(\frac{i}{H/r}\right)^3\right.\\
    &\left.+0.14\left(\frac{e}{H/r}\right)^2\left(\frac{i}{H/r}\right)\right]\text{,}\\
    \end{split}
\end{equation}
where both formulae include the typical type-I damping timescale $\tau_\mathrm{wave}$ derived by \cite{Tanaka2004};
\begin{equation}
    \tau_\mathrm{wave} = \left(\frac{M_{\odot}}{m_\mathrm{pl}}\right) \left(\frac{M_\odot}{\Sigma_\mathrm{gas,pl} a_\mathrm{pl}^2}\right) h_\mathrm{pl}^4 \Omega_{K,\mathrm{pl}}^{-1},\label{eq:tauwave}
\end{equation}
and all quantities are evaluated at the location of the planet, as indicated by the label ``pl''. 
When the embedded planet has carved a gap (Eq.~\ref{eq:crida2006}), we used a type-II damping prescription, where the eccentricity and inclination damping timescales are fixed to $\tau_{\mathrm{mig, II}}/K_\mathrm{damp}$, with $K_\mathrm{damp}=5$, motivated by the fact that damping is more efficient than migration \citep{Lee,Bitsch13,Bitsch20}. 

\subsection{Simulation setup}

From analyses of dust gaps and gas kinematics in the disk around HD 163296, the presence of multiple planets has been inferred \citep{Liu, Zhang2018, Teague2018, Teague19, Rodenkirch}. We chose to model the evolution of the three planet candidates with the most circumstantial evidence of their existence. According to recent results by \citet{Rodenkirch}, the inner planet has an approximate mass of $m_1 = 1.0 \, M_{\mathrm{Jup}}$ and is located at $a_1 = 48\,$AU. For the outer two planets we follow \citet{Teague2018} who found two planet candidates by comparing measurements of rotation curves of CO isotopologue emission with hydrodynamic simulations. The derived semi-major axes of the planets are $a_2 = 83\,$AU and $a_3 = 137\,$AU with corresponding masses of $m_2 = 1.0 \, M_{\mathrm{jup}}$ and $m_3 = 1.3 \, M_{\mathrm{Jup}}$. As a reference, the initial surface density values at these positions are $\Sigma_{\mathrm{gas},0}(\{a_1,a_2,a_3\}) = \{5.86, 2.68, 1.07\}~ \frac{\mathrm{g}}{\mathrm{cm}^2}$.
To account for the uncertainties in semi-major axes and masses inferred for these planets we drew $a_{1,2,3}$ and $m_{1,2,3}$ from Gaussian distributions centered on the aforementioned nominal values. For semi-major axes and masses we chose standard deviations of 10\% and 50\% of the mean value in agreement with \citet{Teague2018}. In doing so, our initial conditions reflect the spread in planet parameters found in other publications \citep{Liu, Zhang2018, Teague19}.

As was done in \citet{Bitsch19}, the initial inclinations and eccentricities of the planets were uniformly distributed between $0.01^{\circ}$ and $0.05^{\circ}$ and 0.001 and 0.01, respectively. All other orbital angles were chosen randomly between $0^{\circ}$ and $360^{\circ}$. The evolution of the planetary system and interactions with the surrounding gas were modeled until the end of the gas lifetime, itself drawn at random as mentioned earlier. Subsequently, the resulting system was further integrated without dissipation until 100\,Myr to investigate long-term stability. In total, we ran three sets of simulations with increasing $\alpha$ viscosity parameters of $\alpha  = \num{1e-4}, \num{3.16e-4}$, and $\num{1e-3}$. Each set consisted of 100 individual setups and the sets differed only by the viscosity parameter $\alpha$ of the disk, thus allowing us to study how the final configurations depend on this parameter. 

\section{Results}
\label{sec:results}
In this section, we first present in detail the results of two selected individual systems that are representative of the general phenomenology, before discussing the ensemble statistics of all simulated planetary systems. We then focus on mass, semi-major axis and eccentricity distributions of the simulated planet population and compare them to the distributions of known exoplanets. To allow for a fair comparison, we only include exoplanets around stars with luminosities exceeding $5 L_{\odot}$ in the statistics, corresponding to the luminosity limit of A-type stars and an approximate lower mass boundary of $1.5\,M_{\odot}$. This way we take the observed correlation of giant planet occurrence rates with stellar mass \citep{Johnson2010} into account and ensure that the host stars have similarly sized inner dead zones as HD 163296 \citep{Flock}. 

\subsection{Individual systems}
\label{sec:individual_systems}
\begin{figure*}[h]
\centering
\begin{subfigure}{.33\textwidth}
    \centering
    \includegraphics[width=\textwidth]{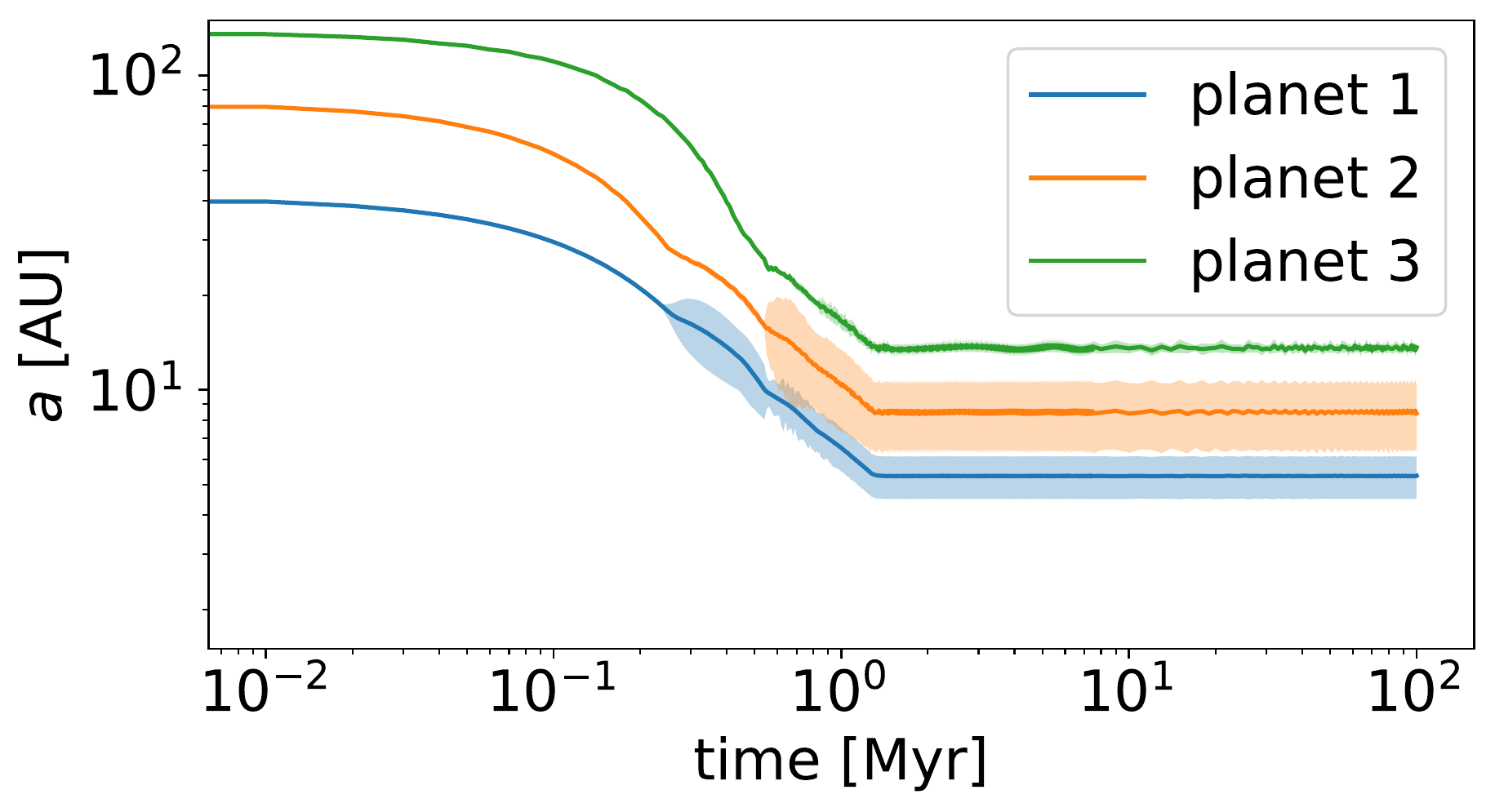}
\end{subfigure}%
\begin{subfigure}{.33\textwidth}
    \centering
    \includegraphics[width=\textwidth]{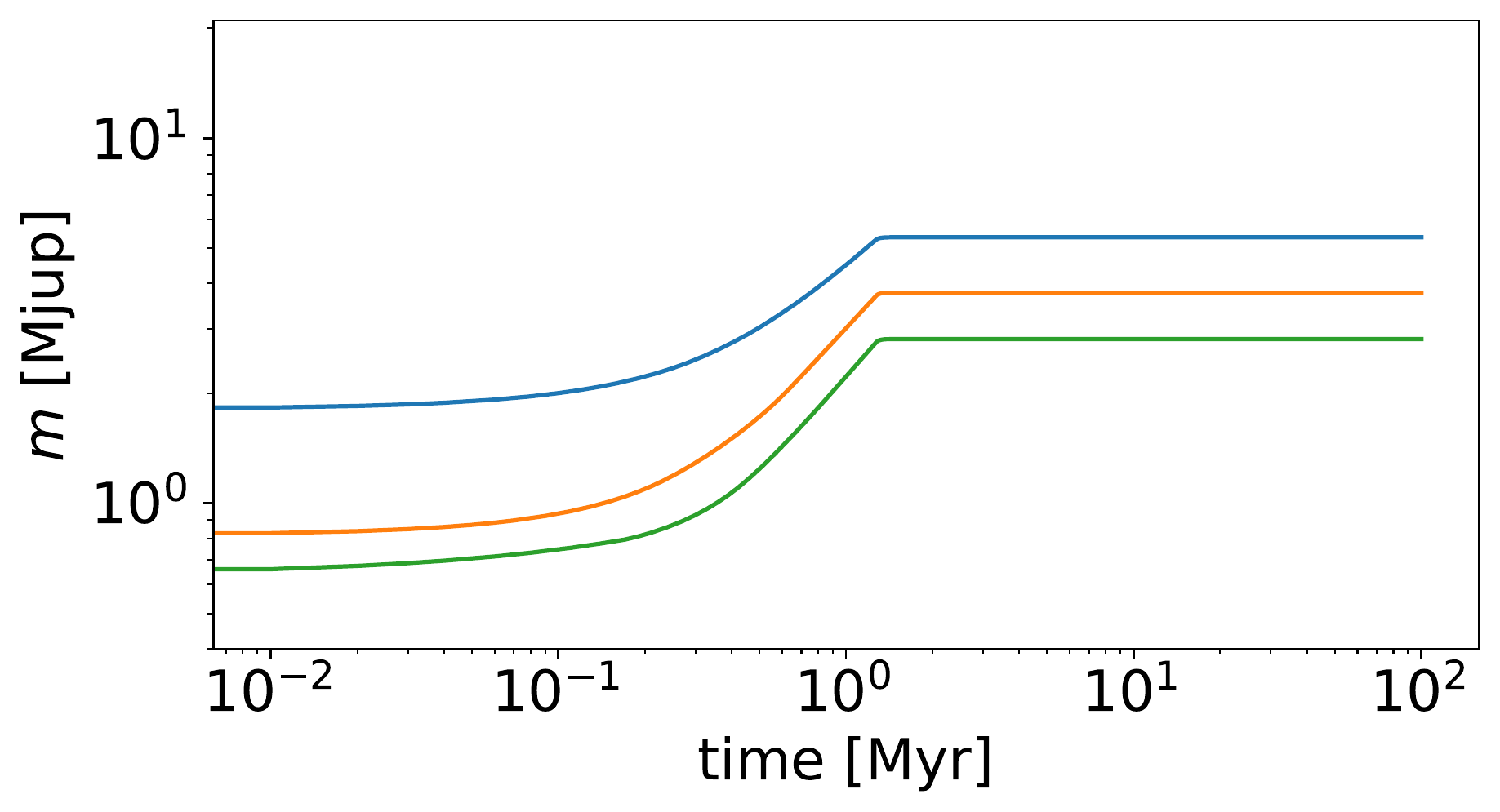}
\end{subfigure}%
\begin{subfigure}{.33\textwidth}
    \centering
    \includegraphics[width=\textwidth]{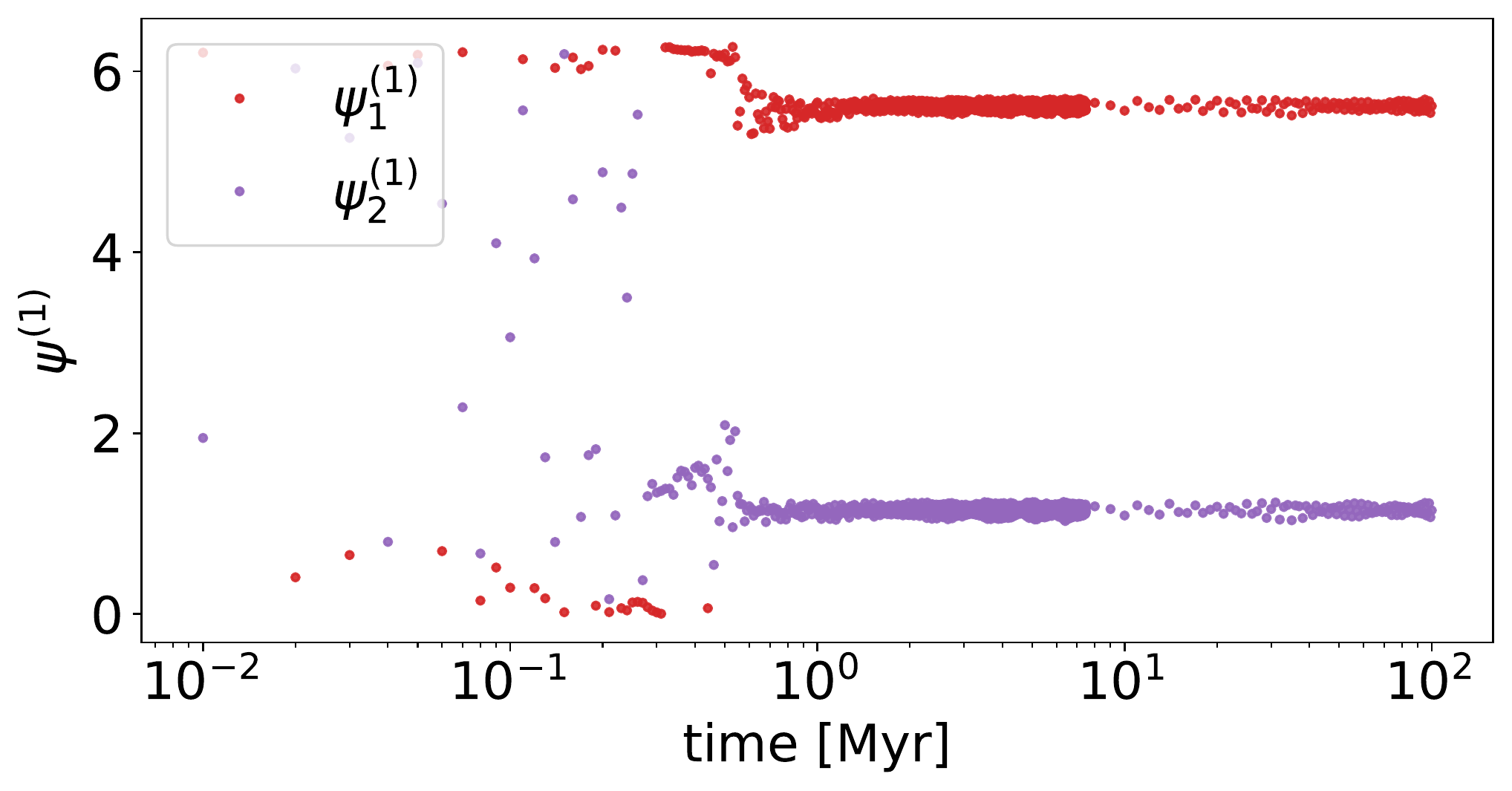}
\end{subfigure}
\begin{subfigure}{.33\textwidth}
    \centering
    \includegraphics[width=\textwidth]{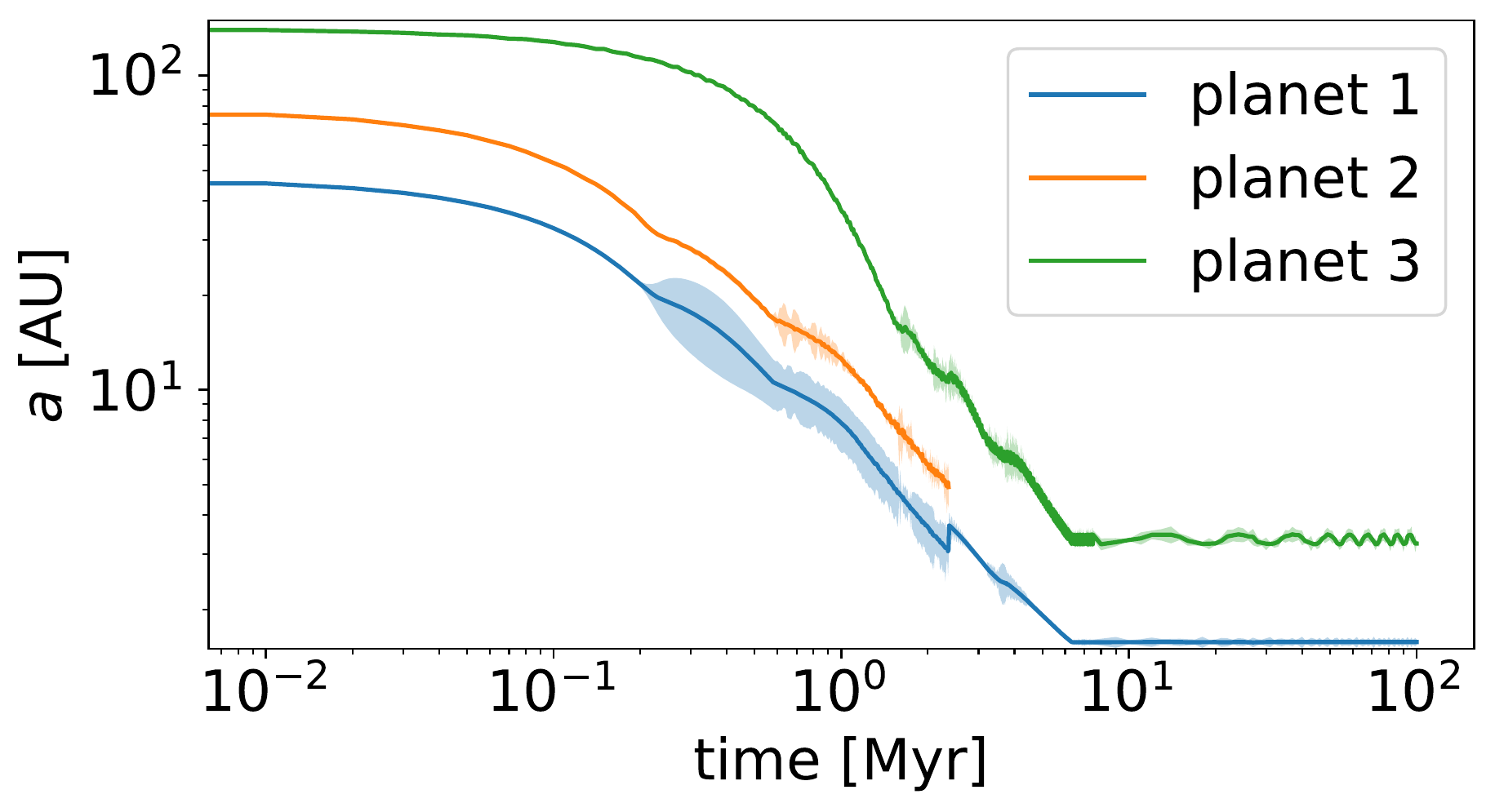}
\end{subfigure}%
\begin{subfigure}{.33\textwidth}
    \centering
    \includegraphics[width=\textwidth]{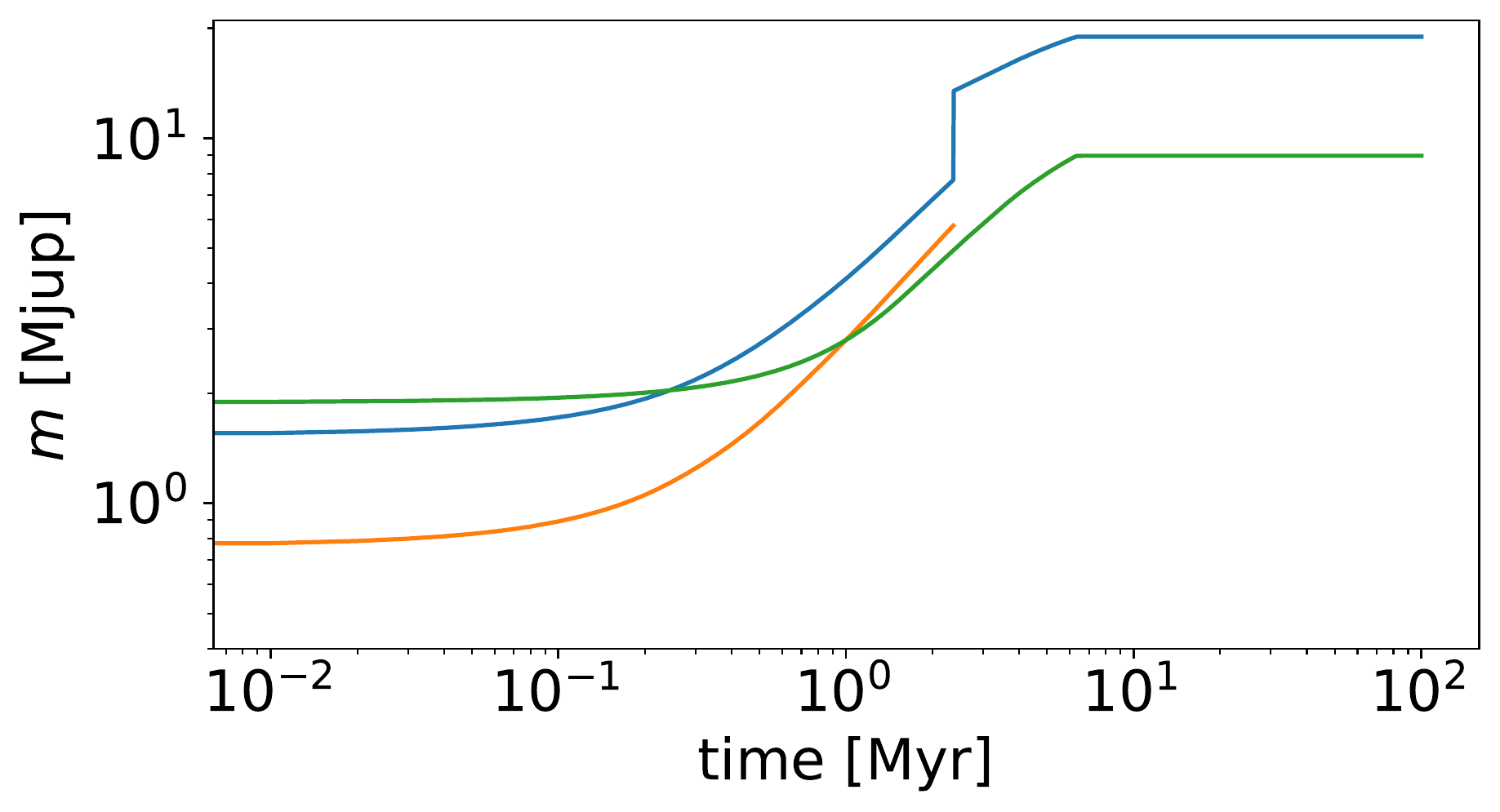}
\end{subfigure}%
\begin{subfigure}{.33\textwidth}
    \centering
    \includegraphics[width=\textwidth]{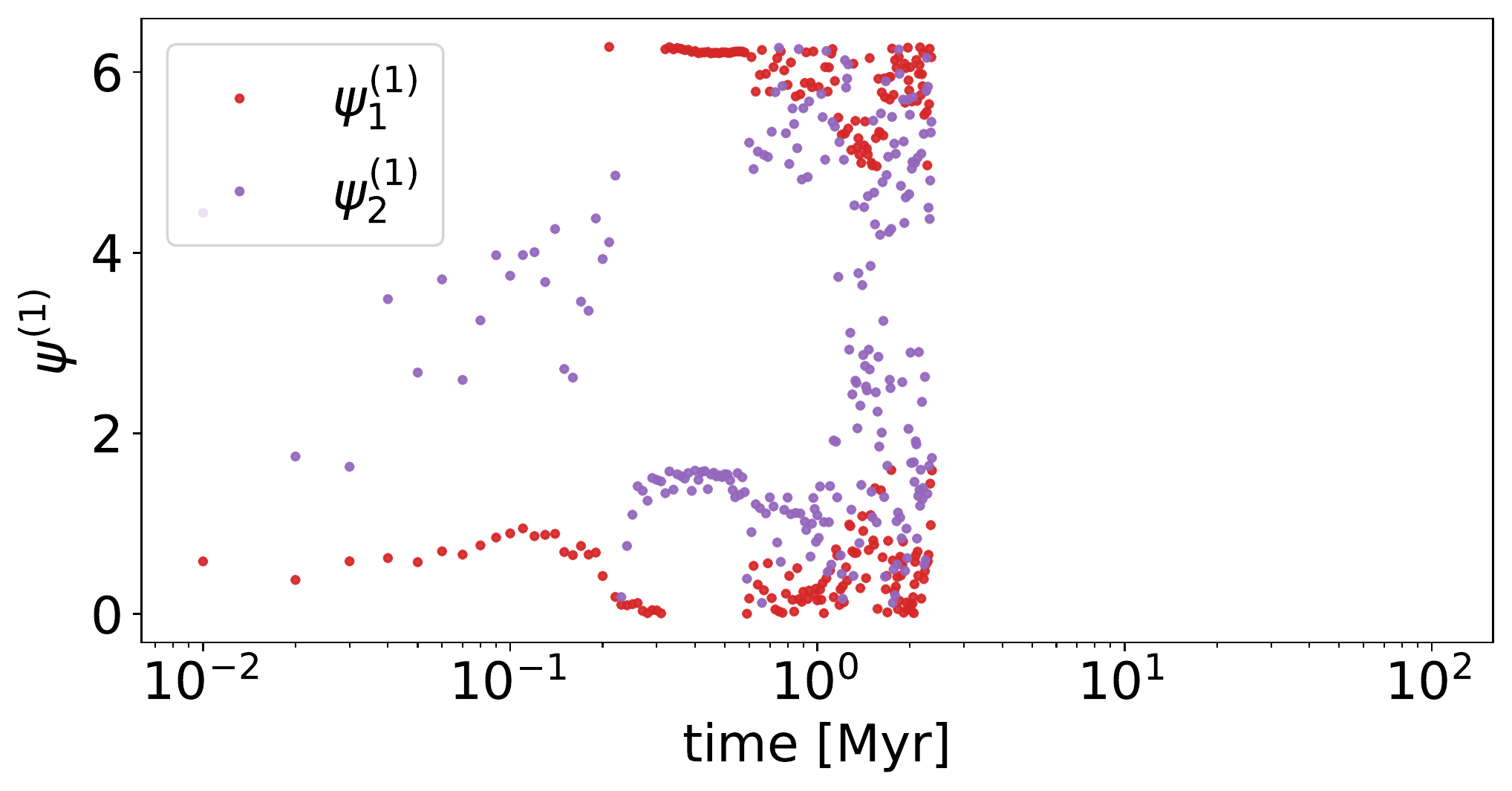}
\end{subfigure}
\begin{subfigure}{.33\textwidth}
    \centering
    \includegraphics[width=\textwidth]{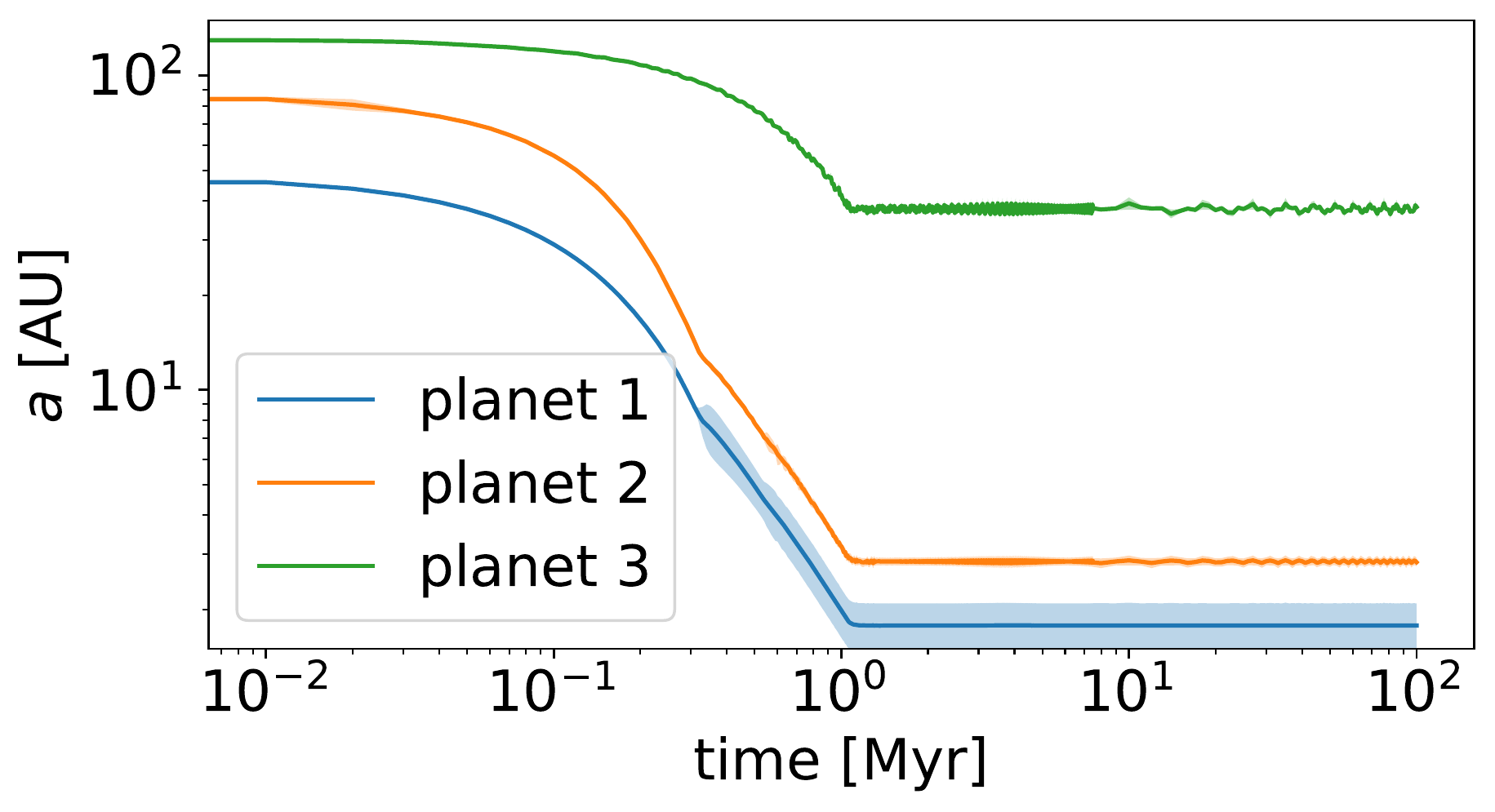}
\end{subfigure}%
\begin{subfigure}{.33\textwidth}
    \centering
    \includegraphics[width=\textwidth]{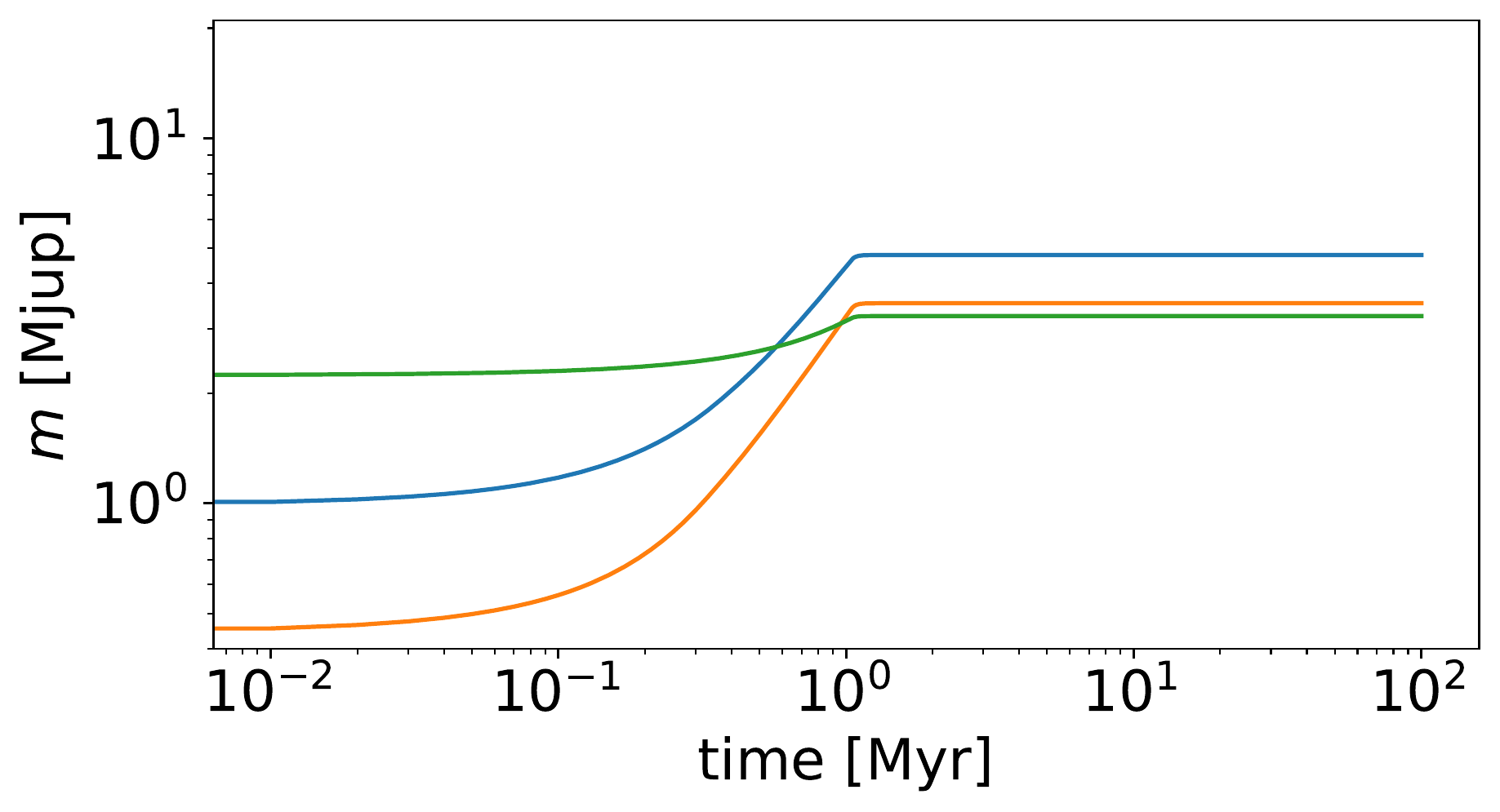}
\end{subfigure}%
\begin{subfigure}{.33\textwidth}
    \centering
    \includegraphics[width=\textwidth]{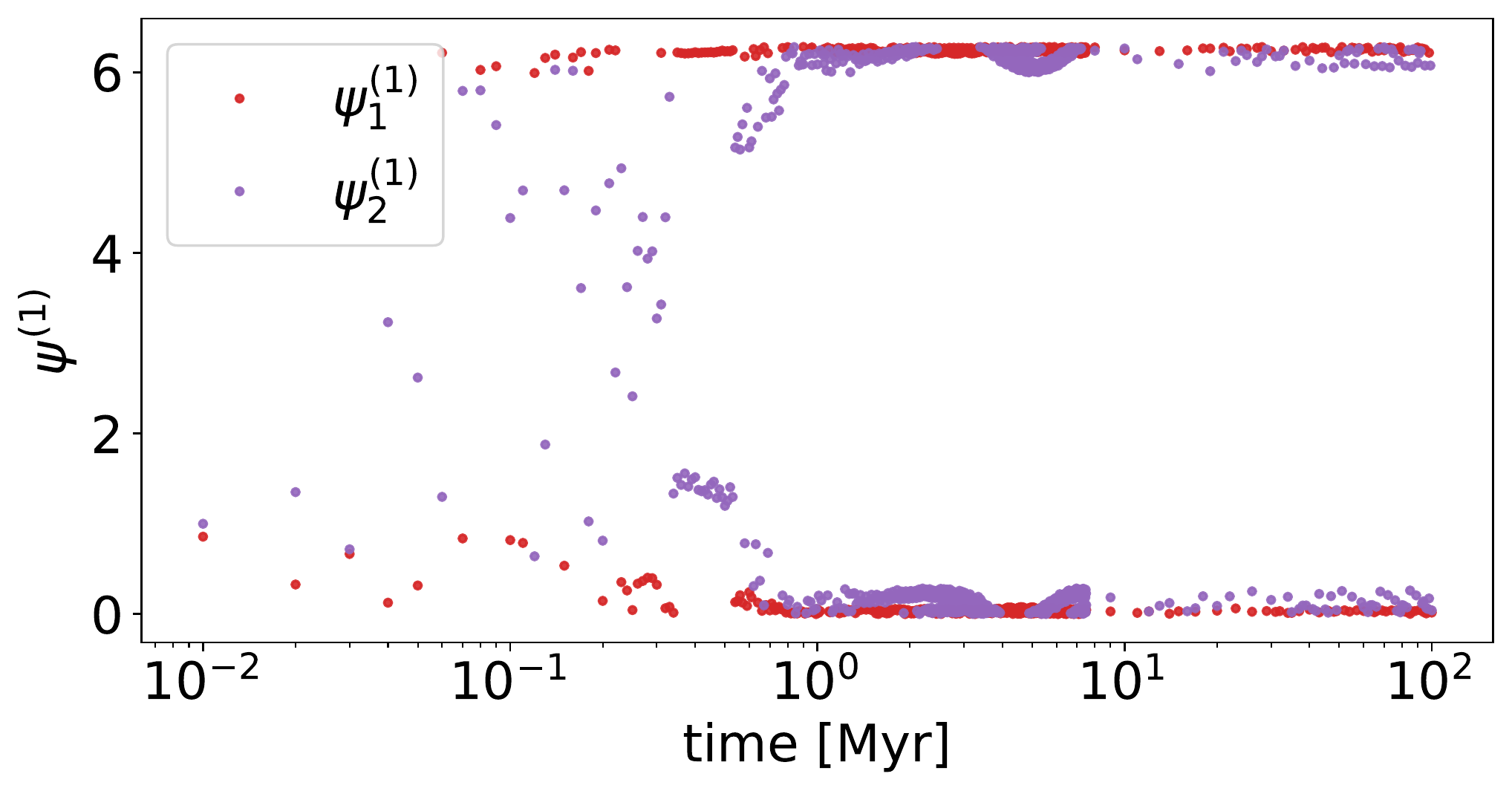}
\end{subfigure}
\caption[short]{Evolution of three selected planetary systems I (top), II (center), and III (bottom) over the integration time of 100\,Myr. The plots display the development over time of the semi-major axes $a_i$ (left), the planetary masses $m_i$ (center) and the resonance angles $\psi^1_1$, $\psi^1_2$ of the inner planet pair (right). The disks in all three systems have $\alpha  = 10^{-3}$. System I and III have gas disk lifetimes of 1.56\,Myr and 1.34\,Myr, respectively, and remain stable throughout the integration whereas system II has a gas disk lifetime of 6.59\,Myr and becomes unstable at 2.36\,Myr when the inner two planets collide and merge.}
\label{fig:individual_systems}
\end{figure*}

It can be insightful to examine the evolution of individual systems in order to interpret the over-all statistics of all simulations. Therefore, in this section we illustrate the effects of orbital migration and gas accretion as well as resonant capture using the example of three selected systems. 

Figure \ref{fig:individual_systems} displays the growth and dynamical evolution of three synthetic systems from here on referred to as systems I, II, and III. In all systems $\alpha$ is equal to $10^{-3}$. All planets migrate inward where the rate of migration depends on the planetary mass and the aspect ratio of the disk at the position of the planet, see Eqs. \eqref{eq:migration_timescale} and \eqref{eq:density_ratio}. Migration stops either with the arrival of the planet at the migration trap at the inner edge of the disk, or at the end of disk life time, which for systems I, II and III occurs at 1.56, 1.34 and 6.59\,Myr, respectively. In all three simulations the initial mass of planet 2 is smaller than for planet 1. Hence, planet 2 migrates faster and the period ratio of the two planets decreases over time, a process known as convergent migration. As the orbits approach each other, the gravitational interaction between the planets drives an increase of their eccentricities (not directly shown but indicated by the shaded region around the semi-major axes which extends form pericenter ($a_\mathrm{pl}(1-e)$) to apocenter ($a_\mathrm{pl}(1+e)$)). 

All the while, the planets grow more massive by accreting gas from the surrounding disk. In the beginning, the rate of gas accretion onto the planets increases because the planets migrate inward toward higher surface densities. At later stages during the disk lifetime, the exponential decrease in surface density becomes noticeable in that the accretion process slows down, which can be seen in the mass evolution of system II. Interestingly, in all exemplary systems the inner two planets enter a 2:1 mean motion resonance within the first Myr. This can be seen from the libration of the mean motion resonance angles $\psi^1_1$ and $\psi^1_2$ which for a general $k$:$(k-1)$ mean motion resonance are defined as
\begin{align}
    \psi^1_1 &= k  \lambda_2 - (k-1) \lambda_1 - \varpi_1\\
     \psi^1_2 &= k  \lambda_2 - (k-1) \lambda_1 - \varpi_2\,.
\end{align}
Here $\varpi_i$ denotes the longitude of the pericenter (which is given by the sum of the longitude of the ascending node and the argument of pericenter of planet $i \in {1,2,3}$), $\lambda_i$ is the mean longitude of the orbit of planet $i$ (defined as the sum of the longitude of pericenter and the mean anomaly) and $k$ is the resonance index, which equals 2 in the case of a 2:1 resonance. In the case of a mean motion resonance the resonance angles librate around a constant value. 

Mean motion resonances are quite common among all simulations (see Sect. \ref{sec:period-ratios}) and can occur both for the inner and outer planet pair. In fact, in system I, not only the inner planet pair experiences resonant motion but the outer pair is locked in a 2:1 mean motion resonance as well. This resonant chain is what prevents the closely spaced system from becoming unstable. In contrast, in system II, the inward migration of planet 3 disturbs the inner two planets which forces the pair out of mean motion resonance and results in the collision of planet 1 and 2 before the end of disk lifetime. However, the rare occurrences of resonant chains are not the only chance for the inner planet pair to remain in a stable configuration. Instead, more often than not the outer planet has a sufficiently low migration rate in order not to migrate to regions where it could disturb the inner planets. This is the case for the majority of simulations with $\alpha  \in \{\num{1e-4}, \num{3.16e-4}\}$ and for comparably high initial masses of planet 3. System III shows an example of this behavior.

\subsection{Planetary masses and orbital distances}

\begin{figure*}[h]
    \includegraphics[width = \textwidth]{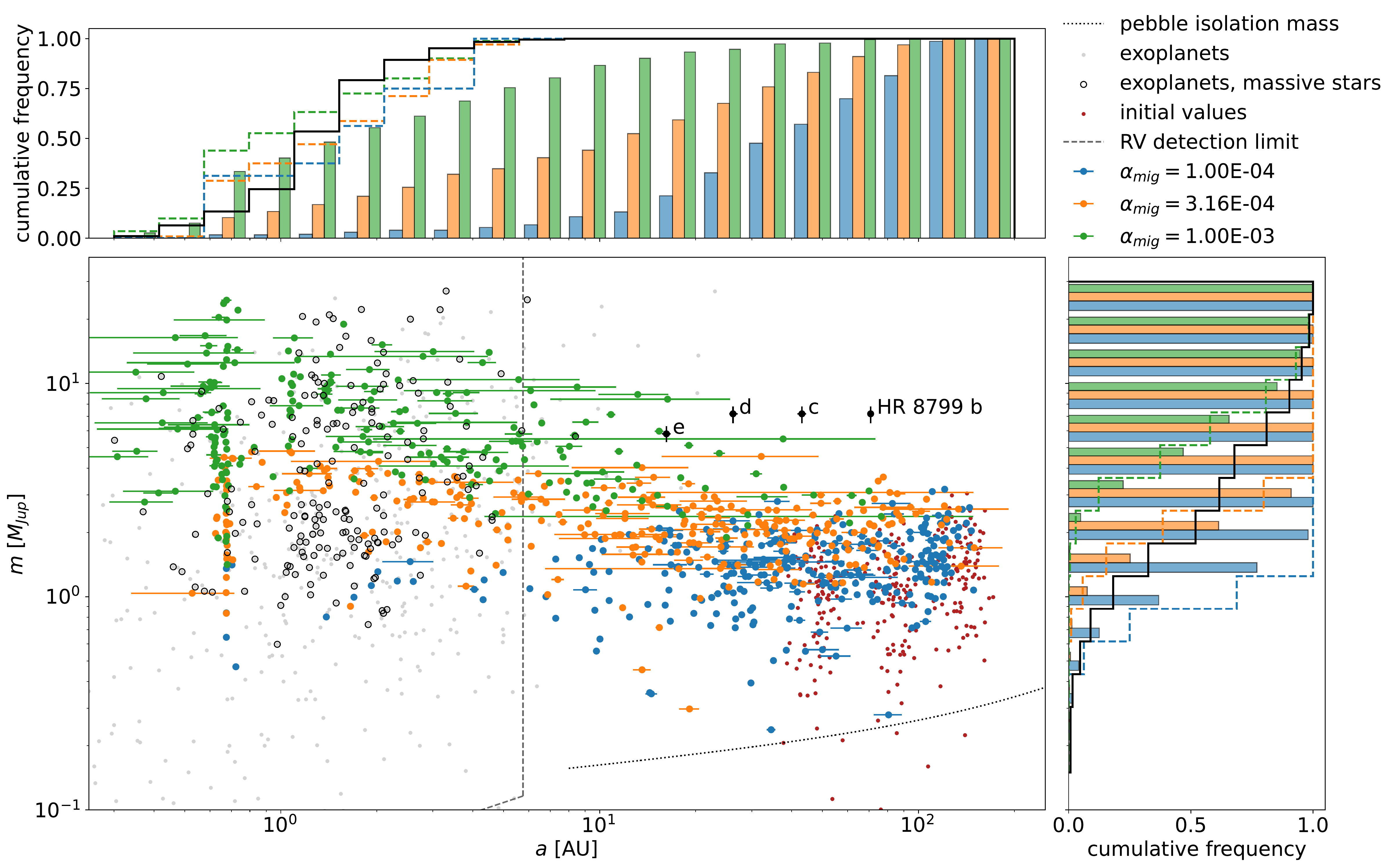}
      \caption{Mass-distance plot comparing all simulated planets to the known exoplanet population. Small red dots indicate the initial values of the simulations. The pebble isolation mass is shown with a dotted line for $\alpha = 10^{-3}$. Blue, orange and green dots denote the planets from simulations with $\alpha$ viscosity parameters $10^{-4}$, $3.16 \times 10^{-4}$, and $10^{-3}$, respectively. The $x$-bars extend from the perihelion to the aphelion distances for a given planet. Gray dots denote the exoplanet population detected via radial velocity and black circles mark those around massive stars with luminosities exceeding $5 L_{\odot}$. For reference, the planets of HR 8799 are shown as black dots \citep{Wang2018}. A dashed line divides the parameter space into a region observable by current radial velocity measurements with a sensitivity of 1\,m/s and observation period of up to 10 years and a currently not observable region. The histograms depict the normalized cumulative frequencies of the final masses (right) and semi-major axes (top). The color scheme is the same as for the main plot, where the bars show the distribution among all simulated planets and the dashed outlines represent the distribution among hypothetically observable planets. The cumulative frequencies among the exoplanets of massive stars with $L > 5 L_{\odot}$ are plotted for comparison with a black solid outline.}
         \label{fig:mass_vs_a}
  \end{figure*}
  
We now turn to the full sample of our synthetic systems. Of particular interest to this study are the final locations of the simulated planets in the mass vs. semi-major-axis plane, especially in relation to the known exoplanet population. Figure \ref{fig:mass_vs_a} shows a superposition of all 300 individual systems. We plot the final locations of all planets that survived the 100\,Myr integration time. The lines across the planet span the range from $a_\mathrm{pl}(1-e)$ to $a_\mathrm{pl}(1+e)$ with $e$ the mean eccentricity averaged over the final 10\,Myr. 

We note that the observed gaps in HD 163296 indicate that the initial planet masses should exceed the pebble isolation mass at the respective positions. For reference, in Fig. \ref{fig:mass_vs_a} we show the estimated pebble isolation mass (Eq. \eqref{eq:Miso}) for $\alpha = 10^{-3}$ as a dotted line. The initial conditions of all simulations are indicated in Fig. \ref{fig:mass_vs_a} as red dots. Almost all simulations start out above $M_{\mathrm{iso}}$ and the few exceptions reach pebble isolation mass within the first Myr of the integration.

\begin{figure}[h]
    \includegraphics[width = \hsize]{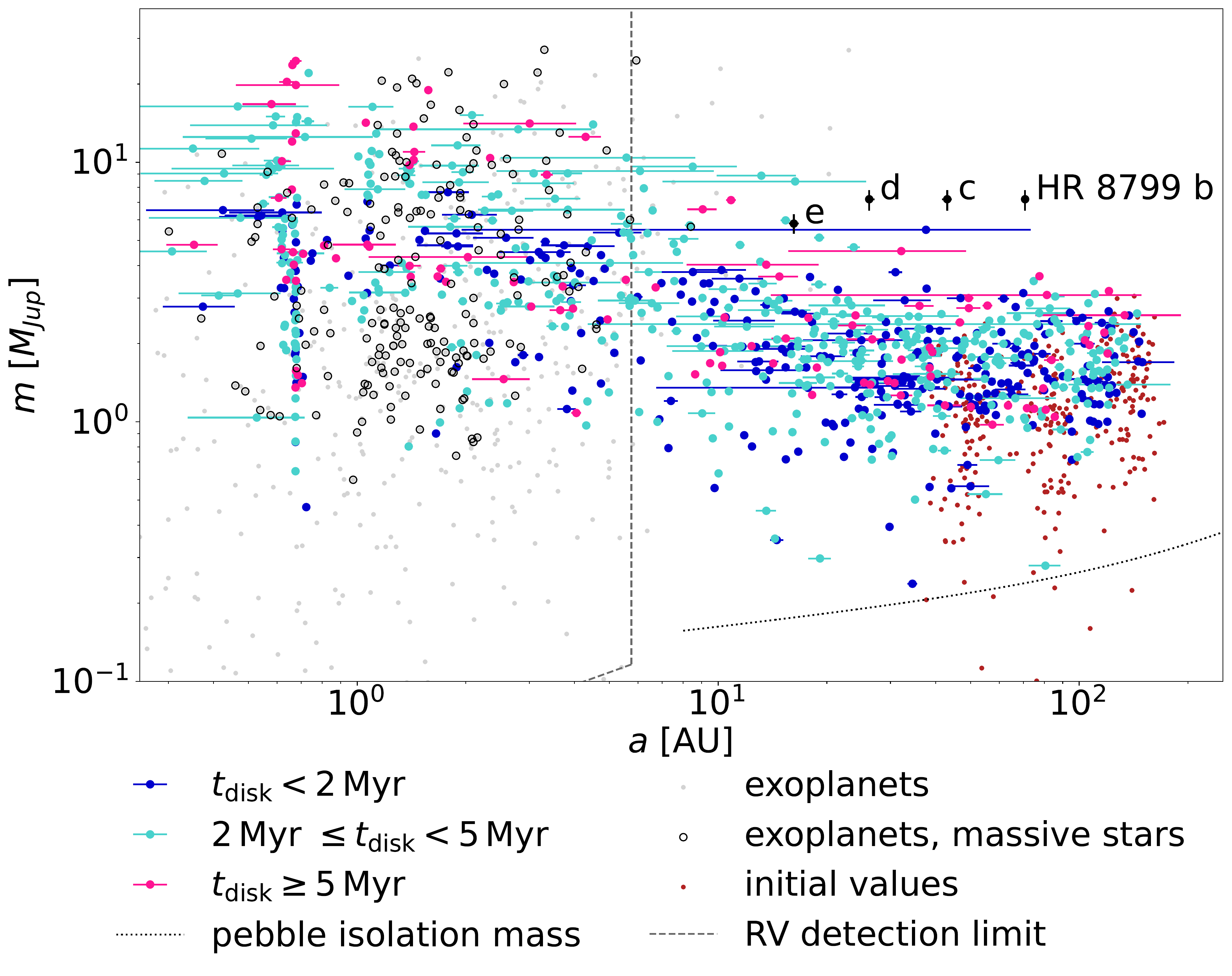}
      \caption{Mass vs.\ semi-major axis plot comparing all simulated planets to the known exoplanet population. The plot is in all aspects the same as Fig.~\ref{fig:mass_vs_a}, except that the simulated planets are colored to reflect the gas lifetime $t_{\mathrm{disk}}$ of the disk in which they formed instead of the $\alpha$ viscosity parameter. Simulated planets with $t_{\mathrm{disk}} < 2\,$Myr, $t_{\mathrm{disk}}$ between 2 and 5\,Myr, and $t_{\mathrm{disk}} \geq 5\,$Myr are shown as colored dots in blue, turquoise and pink, respectively. It is apparent that the dependence of the final orbital parameters on the disk lifetime $t_{\mathrm{disk}}$ is small compared to the dependence on the viscosity parameter $\alpha$ (see Fig.~\ref{fig:mass_vs_a}).}
        \label{fig:mass_vs_a-gaslife}
  \end{figure}

From there, the same evolutionary pattern is evident for all simulations, with rapid gas accretion and orbital migration leading to mass growth and an overall decrease in semi-major axis, visible as a leftward and upward motion in the $m$-$a$-plane. Despite this general pattern, there is a large variance among the final positions and the mass and semi-major axis values cover a range of more than two orders of magnitude. The final masses and semi-major axes correlate to some extent with the gas lifetime of the disk, visualized in Fig.~\ref{fig:mass_vs_a-gaslife}, as well as with the initial positions of the planets. Figure \ref{fig:mass_vs_a-gaslife} shows that a disk lifetime of well over 2\,Myr in addition to the current system age is required for the formation of planets with masses greater than $10\,M_\mathrm{Jup}$. This is in rough agreement with the results of \citet{Schlaufman} who suggest that for planets with masses $>10\,M_\mathrm{Jup}$ gravitational instabilities might be the predominant formation pathway. The dependencies of the final configurations on disk lifetime and initial positions are to be expected because for a given mass a closer initial position of a planet with respect to the star results in a smaller final orbital distance and longer gas lifetimes allow more time for the planets to migrate inward (see also Fig.~\ref{fig:individual_systems}). However, it turns out that the crucial factor in determining the end-positions of the systems is the $\alpha$ viscosity parameter of the underlying disk structure. It is apparent from Fig. \ref{fig:mass_vs_a} that the simulations with $\alpha = 10^{-3}$ constitute the most massive and close-in population. A clustering of planets near $0.7$\,AU can be observed due to the migration trap at the inner edge of the disk. As a consequence of planet-planet interaction, several planets among the $\alpha = 10^{-3}$ set have found their way even further in, to radii as small as 0.3\,AU. To compare with the results of \citet{Bitsch20}, for this choice for $\alpha$ the mean planetary mass is $(6.8 \pm 4.2)\,M_\mathrm{Jup}$, the mean semi-major axis is $(6 \pm 12)\,\mathrm{AU}$ and the mean separation of planets in mutual Hill radii is $8.7 \pm 4.5$.

In contrast, the simulations with the lowest viscosity, corresponding to $\alpha = 10^{-4}$, have lower migration and accretion rates. The planets in this set have final masses generally below $3\,M_{\mathrm{Jup}}$ with a mean mass of $(1.4 \pm 0.5)\,M_\mathrm{Jup}$. The majority remains outside the inner region of 10\,AU and the mean semi-major axis is $(55 \pm 40)\,\mathrm{AU}$. In this case, the mean separation in mutual Hill radii is with $11.1 \pm 5.5$ much larger than for the high viscosity simulations which contributes to the comparably low rate of planet-planet interaction in this set, see Sect. \ref{sec:multiplicities_eccentricities}.

For the ensemble with intermediate viscosity, $\alpha = 3.16 \times 10^{-4}$, a large fraction of planets remain outside 10\,AU but the set also contains several systems in which the inner planet migrates up to the disk edge, thus linking the high- and low-$\alpha$ simulations nearly seamlessly in terms of semi-major axis. The masses of the planets in this set are in between those of the high- and low-$\alpha$ set with mean mass and semi-major axis values of $(2.4 \pm 0.9)\,M_\mathrm{Jup}$ and $(26 \pm 31)\,\mathrm{AU}$. The mean separation of the planets in this set is $12.0 \pm 5.7$ in units of mutual Hill radii. 

To put our results into perspective, we analyze the synthesized giant planet population in the context of the known exoplanets. To this end we use the exoplanet data provided by the NASA Exoplanet Archive as of 03 November 2021. We note that we are not aiming at directly linking the current exoplanets census with our synthetic population of planets, since we are dealing with a single system whose intrinsic properties cannot be generalized to the full sample. Instead, we use the known exoplanet catalog as a gauge to compare our results to the orbital features of known exoplanets. Out of the sample of known exoplanets we select those for comparison to our simulations that have known $m_\mathrm{pl} \sin{(i)}$ values from radial velocity (RV) measurements and orbit stars comparable to HD 163296 with luminosities exceeding $5 L_{\odot}$. For the purpose of this paper we consider all simulated planets as detectable that induce RV amplitudes of more than 1\,m/s and have orbital periods $P<10\,$yrs. This results in a division of the $m$-$a$ plane into an observable and an RV unobservable region as visualized in Fig. \ref{fig:mass_vs_a} by a dashed black line.

Several conclusions can be drawn from the distribution of semi-major axes and masses of the simulated planets. If we assume that embedded planets such as the presumed planets in the HD 163296 disk evolve to form part of the observed exoplanet population, then higher viscosities with $\alpha \gtrsim 3.16 \times 10^{-4}$ are likely required for sufficiently high migration rates. The lower viscosity simulations with $\alpha = 10^{-4}$ seem not to produce enough close-in planets to account for the abundance of giant planets detected with RV measurements. Increased sensitivity of RV and direct imaging observations could help to further constrain the value of $\alpha$, under the assumption that the initial population of wide orbit giants is the norm rather than the exception. If future surveys will uncover a larger number of planets in the region of $m$-$a$ parameter space with $m_\mathrm{pl} \lesssim 3\, M_{\mathrm{Jup}}$ and $a_\mathrm{pl} \gtrsim 10\,$AU, this would point toward an intermediate $\alpha$ viscosity parameter of $\alpha < 10^{-3}$. Instead, if Jupiter-mass planets with semi-major axes of several tens of AU will be ruled out by observations, then our results would imply that the viscosity should be even higher, that is $\alpha \approx 10^{-3}$.

It is apparent in Fig.~\ref{fig:mass_vs_a} that we do not fully reproduce the observed distributions of exoplanet parameters. For example, in the observed population of exoplanets, there is a cluster of planets with semi-major axes between 1 and 3\,AU and masses of 1 to 3$\,M_{\mathrm{Jup}}$ which does not appear in the simulated planet population regardless of the value of the viscous alpha parameter. However, we would like to emphasize that we do not expect to reproduce the observed distribution because our simulations are based on the example of a single disk, albeit with varied initial conditions and disk parameters. Instead, we make statements about the choice for the disk parameter values for which the simulations fit into the general trends of the observed planets.

The best-studied and classic example for direct imaging campaigns to date is the planetary system of the A5V star HR 8799, which provides a benchmark for planet formation models and is one of few discovered systems with multiple directly imaged planets \citep{Marois2008,Marois2010,Wang2018}. It is, therefore, interesting to understand whether the planets of HR 8799 could have formed from rings and gaps as we simulate in our work. When trying to answer this question, we have to keep in mind that the results of these simulations rely on the idea that the gaps and rings observed in protoplanetary disks are caused by planets, an assumption that is still under debate.

The approximate positions of the four directly imaged planets of HR 8799 in the $m$-$a$-plane are shown in Fig. \ref{fig:mass_vs_a} in relation to our simulated planet population \citep{Wang2018}. The semi-major axes of the four planets cover a range typical of our simulations with low viscosity, that is $10^{-4} \leq \alpha \leq 3.16 \times 10^{-4}$. The masses of the planets, derived from luminosity and dynamical constraints, are, however, more in line with our high viscosity simulations with $\alpha = 10^{-3}$. 
The apparent incompatibility between masses and semi-major axes could indicate a mismatch between migration and accretion rates but could also be explained if the planets formed on wider orbits. Alternatively, trapping mechanisms located outside the inner edge could have halted migration. For example, a large number of magneto-hydrodynamic simulations have shown that magnetorotational turbulence in disks can evoke radial structuring in the form of zonal flows \citep[e.g.,][]{Balbus1991, Flock2015}. \citet{Coleman2016} have demonstrated that a radial structuring of the disk can in turn lead to the formation of persistent planet traps at large orbital radii. It is also conceivable that the local gas surface density of the protoplanetary disk might have been higher which would have lead to higher accretion rates.
Since the high masses and large orbits of the HR 8799 planetary system still pose a challenge for core accretion models, gravitational instabilities of the gas disk cannot be ruled out as an alternative formation mechanism either \citep{Boss2011}.
We highlight again that these comparisons are based on the underlying assumption that the HD 163296 disk is representative of the full disk sample which is not known with certainty.

\subsection{Multiplicities and eccentricity distribution}
\label{sec:multiplicities_eccentricities}

For the duration of the disk lifetime the orbital eccentricities and inclinations are tidally damped (see Sect. \ref{sec:migration}). With time, as the disk evolves, the eccentricity and inclination damping become less efficient because of the exponential decrease in surface density. After the dispersion of the disk during the final stages of disk lifetime, the dampening of eccentricities and inclinations ceases altogether. On the other hand, eccentricities and inclinations may increase due to the interaction between multiple planets and the exchange of angular momentum between them. For the observed exoplanets, the eccentricity distribution was found to be consistent with being an outcome of planet-planet scattering (for a review see \citet{Davies} and references therein). However, in the absence of other sources to disturb the system, these simulations require a minimum number of three planets to trigger sufficient scattering events \citep{Weidenschilling, Juric2008, Raymond2009,Sotiriadis2017}.

\begin{figure}[h]
   \centering
   \includegraphics[width=\hsize]{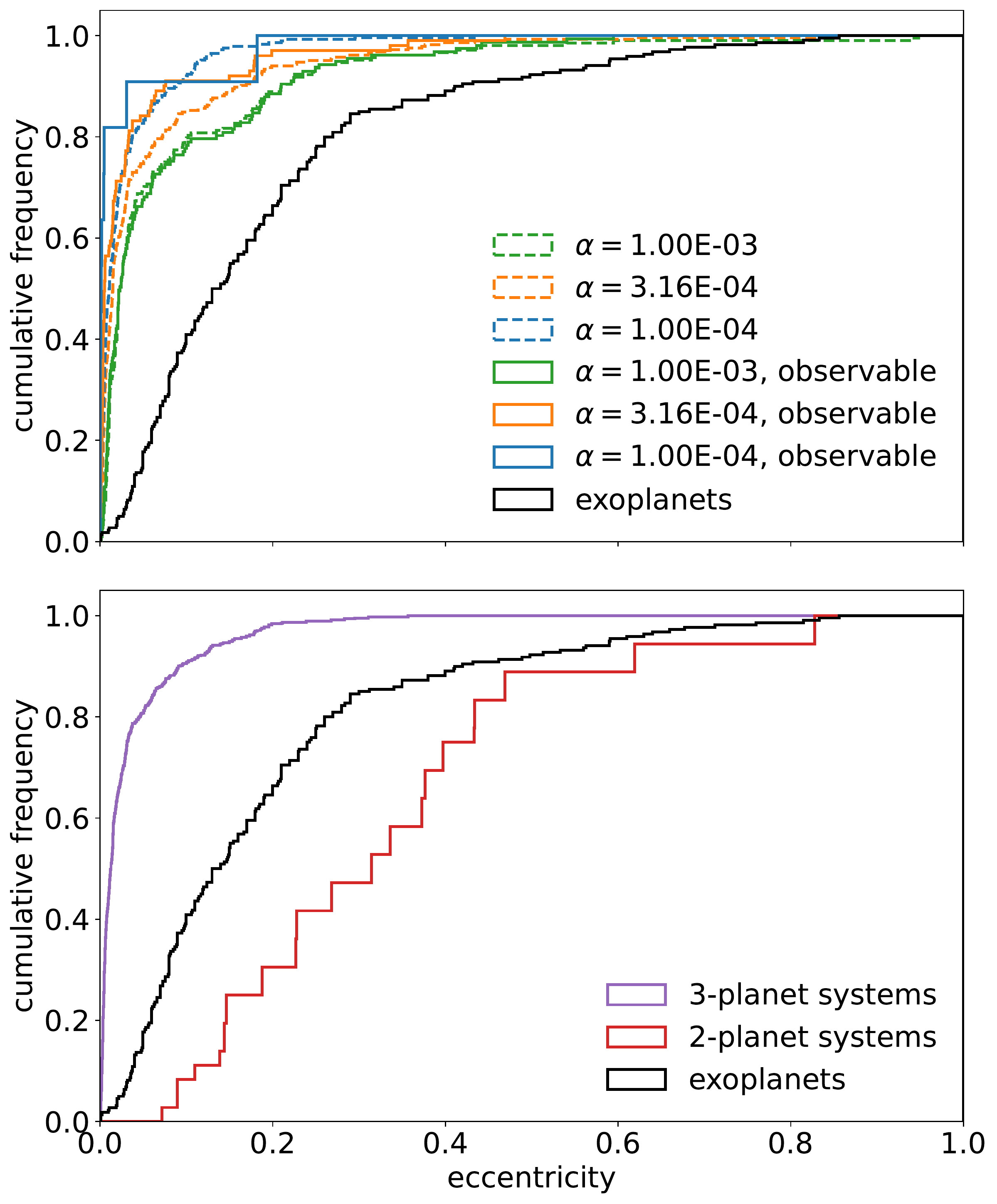}
      \caption{Cumulative histogram of the eccentricity distribution among the simulated planet population. 
      In the upper panel, the normalized distributions of simulations with $\alpha$ viscosity values $10^{-4}$, $3.16 \times 10^{-4}$, and $10^{-3}$, are depicted in blue, orange and green, respectively. Dashed lines indicate the eccentricities among all simulated planets, whereas the eccentricity distributions of only the hypothetically observable planets are shown as solid lines. The lower panel confronts the eccentricity distribution of planets in stable 3-planet systems (purple) with 2-planet systems (red) in which one planet collided or got ejected. For comparison, the eccentricity distribution of observed exoplanets around stars with luminosities exceeding $5 L_{\odot}$ is shown as a black solid line in both panels.}
         \label{fig:eccentricity_hist}
\end{figure}

These mechanisms -- the tidal damping and the effect of planet scattering -- shape the cumulative distribution function of eccentricities of our simulated planets, which we show in Fig. \ref{fig:eccentricity_hist} for different $\alpha$ parameters. The planetary systems with lower $\alpha$ values typically remain in stable configurations where the planetary orbits are widely separated and no close encounters occur. Accordingly, the eccentricities are comparably low, with a mean value of $0.028$ for $\alpha = 10^{-4}$.

For higher viscosities the migration rates increase causing more planets to migrate to the inner edge. Therefore, close encounters occur more frequently and the gravitational interactions among the planets lead to mutual excitation of the eccentricities. This is evident in a significant increase of planet eccentricities with increasing $\alpha$, with mean values of $0.052$ and $0.076$ for simulations with $\alpha = 3.16 \times 10^{-4}$ and $10^{-3}$, respectively. For comparison, the mean of the observed eccentricities among the detected exoplanets, shown in Fig. \ref{fig:eccentricity_hist} to be around $0.135$, is even higher than for our $\alpha = 10^{-3}$ set.

The moderate eccentricities among the simulated population could be a result of the relatively low number and large initial separations of planets in the disk. A higher mean eccentricity could be achieved with additional planets in the synthetic systems which is not unreasonable, considering the growing evidence for a fourth planet in the HD 163296 disk \citep{Pinte, Teague19, Teague2021}. Furthermore, eccentricities could be increased via external perturbations (for instance, by planetesimal belts such as those invoked in the Nice model to drive the giant planet instability in the Solar System, \citep{Raymond2009b}) which can act as catalysts for dynamical excitation. In addition, observations of exoplanets might be slightly biased toward higher eccentricity values. More precisely, recent observations \citep{Anglada, Wittenmyer2013, Kuerster2015, Boisvert2018, Hara2019, Wittenmyer2019} indicate that in cases of sparse data, two giant planets in circular orbits can mimic the radial velocity signature of a single eccentric giant planet, thus, leading to potential misinterpretations of the eccentricity distribution.

For the exoplanet population it was shown that eccentricities correlate inversely with system multiplicities \citep{Bach}. The same is true for the simulated set of planets, as is apparent in the lower panel of Fig. \ref{fig:eccentricity_hist}. In the 2-planet systems the eccentricities of the remaining planets are driven toward higher values simply because these are the systems in which strong mutual interaction occurred which lead to dynamical instabilities and scattering events. Therefore, the 2-planet systems have generally higher eccentricities than their 3-planet counterparts. Given that in the type-II regime the migration timescale scales inversely with the $\alpha$ viscosity parameter, it is not surprising that with 4, 10, and 76 ejections and collisions for the $\alpha = \num{1e-4}, \num{3.16e-4}, \num{1e-3}$ sets, respectively, the majority of scattering events occurs for the set with highest $\alpha$ value, $\alpha = \num{1e-3}$.

The above ejection and collision rates correspond to an average of 2.96, 2.9, and 2.24 surviving planets per system for the $\alpha = \num{1e-4}, \num{3.16e-4}, \num{1e-3}$ sets, respectively. Considering only the hypothetically observable planets, these values are reduced to system multiplicities of 0.16, 1.04 and 1.71. The apparent discrepancy between observed and actual multiplicities implies that in systems observed to harbor one giant planet, several others might still be hidden. This is in line with findings by \citet{Bitsch20} whose simulations indicate that giant planets should, on average, occur in multiples rather than singly.

\subsection{Period ratios}

\begin{figure}[h]
   \centering
   \includegraphics[width=\hsize]{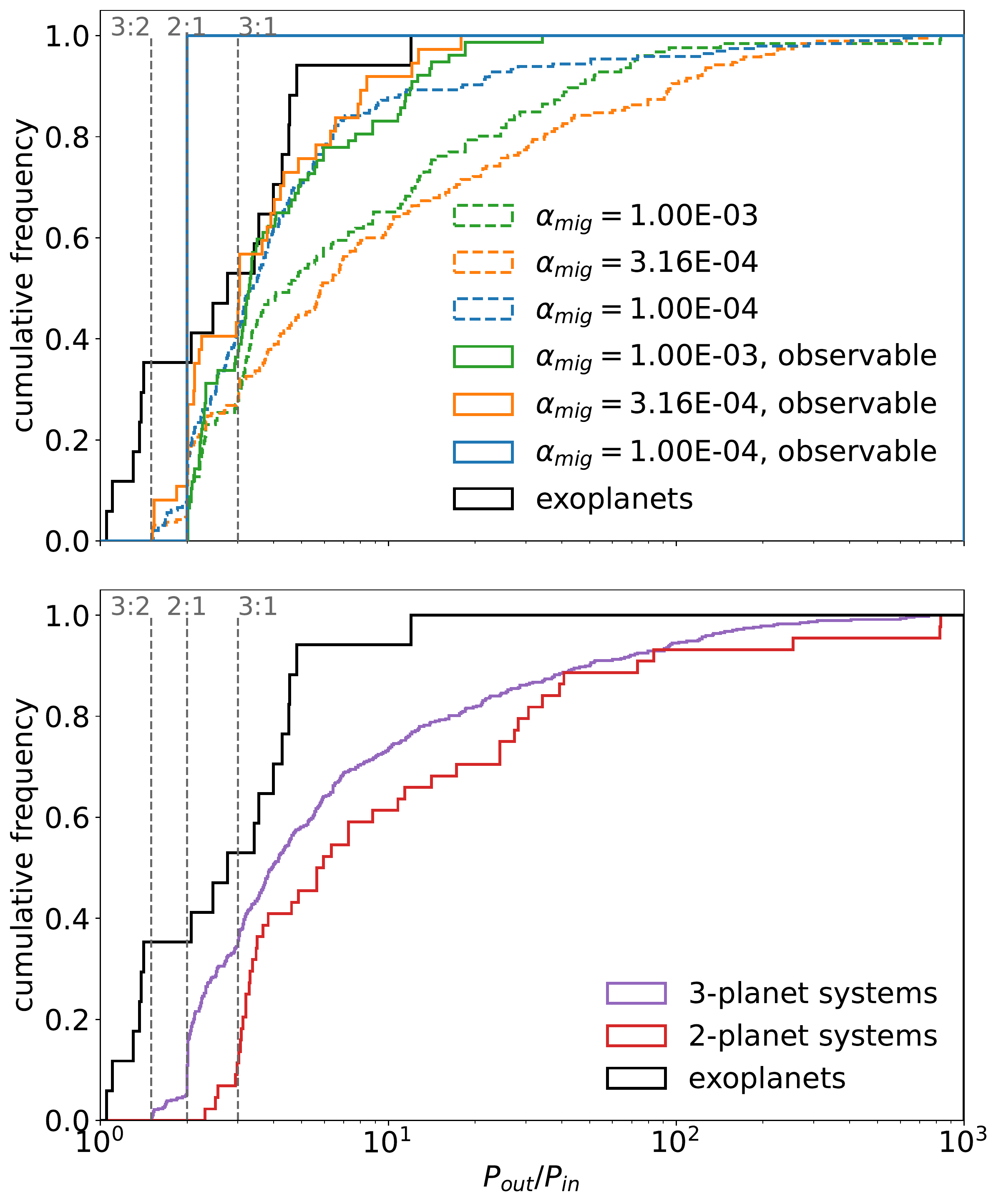}
      \caption{Normalized cumulative distribution function of the period ratios among simulated planet pairs. In the upper panel, the period ratio distributions of simulations with $\alpha$ viscosity values $10^{-4}$, $3.16 \times 10^{-4}$, and $10^{-3}$, are depicted in blue, orange and green, respectively. Dashed lines indicate the period ratios among all adjacent planet pairs, whereas period ratios of those planet pairs that would hypothetically be observable are shown as solid lines. The lower panel compares the period ratio distribution of planets in 2- (purple) and 3-planet systems (red). We observe a significant pile-up of period ratios near the 2:1 resonance for 3-planet systems and around the 3:1 resonance for 2-planet systems.}
         \label{fig:periodratio_hist}
\end{figure}

\begin{figure*}[h]
   \centering
   \includegraphics[width=\hsize]{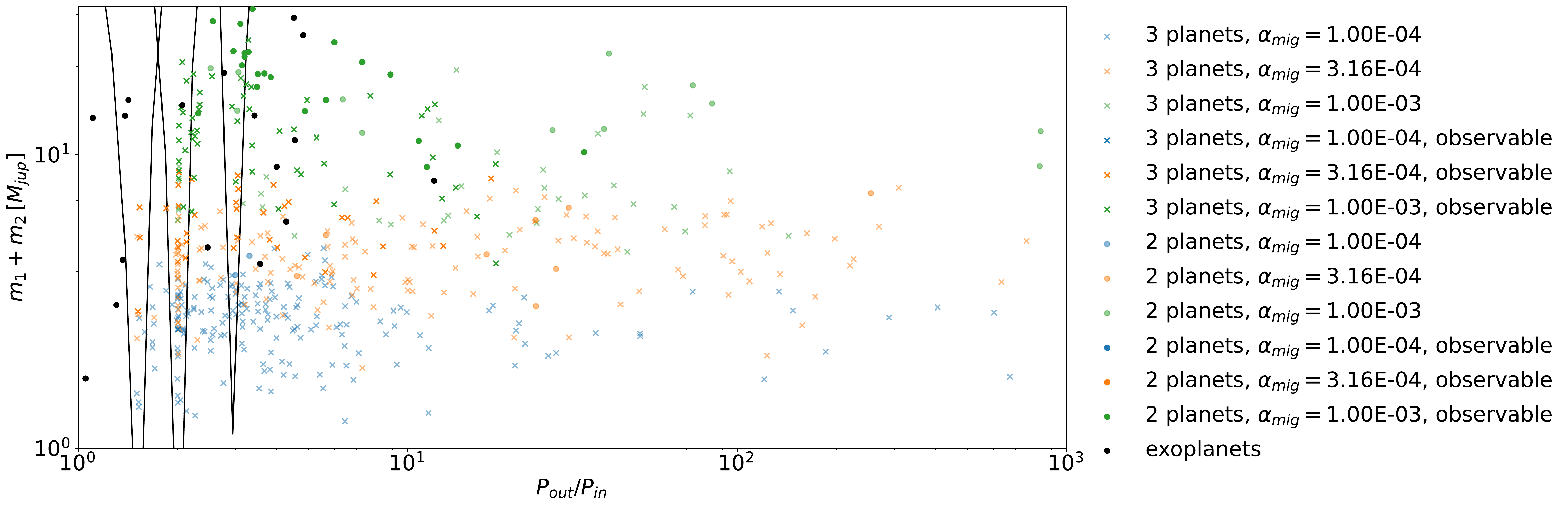}
      \caption{Final period ratios of all simulated planet pairs plotted against their combined mass. Crosses mark planet pairs in 3-planet systems and circles denote 2-planet systems, where the third planet collided or got ejected. The transparency of the markers indicates whether or not the pair would be observable with current standard radial velocity measurements. The different colors denote the $\alpha$ viscosity value of the protoplanetary disk with blue, orange and green for $\alpha = 10^{-4}$, $3.16 \times 10^{-4}$, and $10^{-3}$, respectively. The period ratios of known exoplanet pairs around stars with luminosities exceeding $5 L_{\odot}$ are marked with black dots. The horn-shaped black solid curves indicate the approximate resonance boundaries $\left(\nicefrac{P_{out}}{P_{in}} - \nicefrac{p}{q}\right) = 0.05 \left(\nicefrac{(M_1+M_2)}{M_{Jup}}\right)^{1/2}$ for a p:q mean motion resonance, in this case for 3:2, 2:1, and 3:1 resonances. It is evident that period ratios close to the 2:1 mean motion resonance occur frequently for all $\alpha$ viscosity values. But since for the lower two $\alpha$ viscosity values the distances to the star are typically larger only few of the resonant pairs would be observable.}
         \label{fig:mm_resonances}
\end{figure*}

\label{sec:period-ratios}
The period ratios in multiple planet systems reflect the spacing of neighboring planets and the dynamical configuration of the system. In Fig. \ref{fig:periodratio_hist}, we show the period ratios of adjacent planet pairs from our simulations. The normalized histograms in the top panel depict the cumulative distribution of period ratios for simulations with different $\alpha$ viscosity parameters. Dashed lines indicate the distribution among all simulated planets, the distributions among hypothetically observable planet pairs are shown as solid lines. Since lower $\alpha$ viscosity values imply slower migration, there are fewer planet pairs with $\alpha = 3.16 \times 10^{-4}$ in the observable region than for $\alpha = 1.0 \times 10^{-3}$. This can also be seen from Fig. \ref{fig:mm_resonances} where we show the combined mass of the planet pairs as a function of their period ratio. For the lowest $\alpha$ value, only one planet pair enters the observable region.

Mean motion resonances occur frequently among all our simulations and can arise both for the inner and outer planet pair. Independent of the viscosity parameter, the distributions show an accumulation of period ratios near the 2:1 and 3:1 mean motion resonances. The majority of 2:1 resonances arises for the inner planet pair. In these cases the second planet approaches the inner one either because it is smaller and migrates faster or because the migration of the innermost planet is stopped at the inner edge of the disk while the second planet continues to migrate further in. This resonance can act as a stabilizing factor in cases of closely packed systems. If, however, no resonance occurs or the resonance is disturbed by the third planet, as is the case for system II in Sect. \ref{sec:individual_systems}, the system becomes unstable and at least one planet collides with another planet or the star or it gets ejected.

In the lower panel of Fig. \ref{fig:periodratio_hist} the difference in period ratios between 3-planet and 2-planet systems is shown. Interestingly, there is a pile-up of period ratios near the 3:1 resonance for the remaining planet pair in unstable 2-planet systems.

The clustering of period ratios near the 2:1 and 3:1 resonances that is apparent in Fig. \ref{fig:mm_resonances} is interesting also in the light of recent findings by \citet{Bitsch20}, whose simulations suggest that mean motion resonances should be rare among giant planets. \citet{Bitsch20} argue that the lack of resonances might be due to a low $\alpha = 10^{-4}$ parameter and hence slow migration rates. The fact that resonances occur frequently in our simulations indicates that different formation pathways can result in different system architectures. For instance, the simulation set up used by \citet{Bitsch20} differs from ours in that the number of planetary embryos is much higher, ranging from 15 to 60. At the same time, their systems are more compact and the outer embryos are located at a maximum distance of 17\,AU. Both factors lead to an increase in the number of scattering events in comparison to our simulations.

A direct comparison with observations is currently not feasible because there are too few detected giant planet pairs around luminous stars to draw meaningful conclusions. Moreover, given that the diversity of disks and initial solid distribution is expected to yield different system configuration, we cannot draw strong conclusions from our single-disk simulations. However, considering all known multiplanet systems, there seems to be a clustering of period commensurabilites for planet pairs with a total mass exceeding 1\,M$_{\mathrm{Jup}}$ \citep{Wright, Winn} which would agree with our results. Currently, we are limited to the example of HD 163296 but future discoveries of other protoplanetary disks with strong evidence of multiple planets will help significantly to understand whether the simulated processes can fully reproduce the observed eccentricity and period distributions.

\section{Discussion}
\label{sec:discussion}

In this section, we discuss the underlying assumptions of our modeling and outline some of the implications of our results for exoplanet and protoplanetary disk observations.

\subsection{Effect of changes to the disk model}
The outcome of our simulations is largely determined by the $\alpha$ viscosity parameter of the underlying disk. Since most relevant quantities, such as the migration and accretion rates scale with $\alpha$, see Eq. \eqref{eq:acc_rate3}, and the $\alpha$ parameter values of the simulations span an order of magnitude, we do not expect qualitative changes in our findings from changes in the disk structure by a factor less than that (such as a factor 2 in surface density as proposed by \citealt{Calahan}).

In this work, we limited ourselves to disk models with constant $\alpha$ viscosity parameter as these are the most commonly applied. The effects of a varying $\alpha$ parameter as suggested by \citet{Liu}, for example, and applied by \citet{Rodenkirch} remain to be tested but are not expected to cause significant changes in our results. In the case of the profile used by \citet{Rodenkirch}, for example, the $\alpha$ parameter at the location of the innermost planet is set to $\alpha(48\,\text{AU}) \approx \num{2e-4}$, meaning that the inner planet would likely evolve in a similar manner as the inner planets in our $\alpha = \num{1e-4}$ simulations. The increase of $\alpha$ with radius would cause the outer planets to migrate and grow faster than in our simulations until they reach the inner parts of the disk. Therefore, a probable result of these changes would be a slightly more closely spaced final planet population. The decrease in relative distances would likely lead to an increased level of planet-planet interaction and somewhat higher planet scattering rates with respect to our $\alpha = \num{1e-4}$ simulations. 

\subsection{Constraints for protoplanetary disk parameters}
Our simulations are based on the premise that the supposed planets in HD 163296 actually exist. If, in addition, we assume that such massive planets on wide orbits are typical of young exoplanet systems, at least for host stars of similar sizes, this enables us to draw conclusions about the parameters of protoplanetary disks. More precisely, if the putative planets of HD 163296 are common examples of young exoplanet systems they should evolve so that their final planetary and orbital parameters match those of the discovered exoplanet population. Our models show that the agreement of the final mass and semi-axis distributions of the simulated planets with the observations depends crucially on the viscosity parameter $\alpha$ of the disk and requires that $\alpha \gtrsim 10^{-4}$. From the masses and semi-major axes of the synthesized planets we constrain the viscosity parameter to an approximate range of $\num{3.16e-4} \lessapprox \alpha \lessapprox 10^{-3}$. A $\alpha$ parameter toward the lower end of this range is expected to cause the formation of a larger number of planets with masses below $m_\mathrm{pl} \lesssim 3\,M_\mathrm{Jup}$ and orbits outside $a_\mathrm{pl} \gtrsim 10\,$AU whereas $\alpha \approx 10^{-3}$ results in a more massive and close-in population. This prediction provides a way to further constrain the value of $\alpha$ with future observations. 

Independent of the choice for $\alpha$, many planets remain outside the region in $m$-$a$ parameter space that is accessible to RV observations at current measurement sensitivities. Therefore, the multiplicities of the simulated systems that would hypothetically be observed are far lower than the actual system numbers. Still under the assumption that the HD 163296 planets are common examples of young and massive exoplanets, this result suggests that exoplanet multiplicities are likely being underestimated and that there might be hidden companions in systems known to harbor at least one giant planet. Indications of a potential underestimation of multiplicities due to the RV detection limit were also found in previous studies by \citet{Wagner2019} and \citet{Bitsch20}, among others.

Future direct imaging campaigns are a promising alternative to RV measurements for detecting planets in orbits beyond 10\,AU. With the exception of HR 8799, direct imaging has so far mostly discovered systems with no or single planets. If this trend continues in future observations, it would suggest larger values of $\alpha$ on the order of $\alpha \approx 10^{-3}$, because for higher viscosity the giants should migrate to the inner disk regions where they would no longer be detectable by direct imaging. On the other hand, if the viscosities are low, that is $\num{1e-4} \lessapprox \alpha \lessapprox \num{3.16e-4}$, a larger number of Jupiter-mass planets should be detected at distances of $a>10\,$AU. As for the number of planets per system, the observed low number could also indicate that planets forming in rings similar to our simulations should mostly have only one survivor. These results, of course, all rely on the assumption that the observed gap and ringlike structures in protoplanetary disks are in fact caused by planets, which to date is not entirely clear.

\subsection{Implications of mean motion resonances}
In contrast to \citet{Bitsch20} whose findings indicate that mean motion resonances among giant planets should be rare, (near)resonant configurations of planets occurred frequently in our simulations. We conclude that the cause of this discrepancy is likely to be the difference in formation pathways, see Sect. \ref{sec:period-ratios}. A conceivable implication for observed exoplanet systems is that those systems exhibiting resonances originated from disks with fewer and or more widely spaced planetary embryos. In this way, the chance of scattering events would have been reduced and newly formed resonant configurations would have been less likely to be perturbed by additional scattered or migrating bodies. However, our simulations presented here are not able to reproduce the eccentricity distribution of the giant planets, indicating that a lower number of initial embryos might be responsible for forming systems in resonance, while systems with many more initial embryos could explain the eccentricity distribution of the giant planets. This actually indicates that there are maybe different planet formation pathways for giant planet systems.

In our simulations, there appears to be a connection between the number of hypothetically observable mean motion resonances among the planets and the underlying viscosity parameter $\alpha$ of the simulated disk. Under our assumptions for the formation pathway detailed above, it is therefore conceivable to use the abundance of observed exoplanet pairs in mean motion resonance to narrow down the magnitude of the $\alpha$ viscosity parameter in disks such as the HD 163296 protoplanetary disk. Since for lower $\alpha$ values the distances covered by migration are smaller, fewer exoplanet pairs will exist for which both planets are observable. This implies that fewer resonant systems would be observed whereas the opposite would be true for high $\alpha$ values. A step in this direction has already been taken by \citet{Leon2021}, who derived information about the viscosity and surface density of the former protoplanetary disk from the observed orbital architecture of the Kepler-223 system.

\subsection{Location of rings}
For higher $\alpha$ parameters and, as a result, increased migration rates, one would expect the ring positions to evolve in time as the planets migrate inward, because the ring caused by the planet should follow the planet's inward migration. For similarly sized disks, the location of the outermost ring should, therefore, be a function of the disk age, still under the assumption that all rings are caused by forming planets. So far, no such correlation has been observed \citep{Marel}. This could indicate lower viscosities but might also be a consequence of the relatively small number of resolved disks in the sample. Different migration rates should in principle also be reflected in the morphology of the gaps and rings induced by the planet \citep{Meru, Weber2019}. However, using this approach to identify migrating planets has proven to be difficult and has not yet been tried for the disk of HD 163296 \citep{Nazari}.

\section{Summary and conclusions}
\label{sec:conclusion}

Detection of planets in their early stages of evolution is crucial to test and improve current planet formation theories and to constrain the initial conditions of population synthesis models. To date, however, only one system harboring young planets and a protoplanetary disk has been detected by direct imaging \citep{Keppler} and for few other stars the presence of such embedded planets has been put forth. 

In this paper, we investigated one of these rare cases, namely the planet candidates around HD 163296, inferred from hydrodynamical simulations and the observations of gas kinematics in the protoplanetary disk \citep{Isella+2016, DSHARP-Huang, Isella+2018, Zhang2018, Liu, Teague2018,Pinte, Teague19,  Rodenkirch}. We performed $N$-body simulations that model the evolution and growth of the three putative planets and their interactions with the surrounding disk. In contrast to previous work on this topic, we considered multiplanet systems and the effects of gravitational interaction between planets. Furthermore, we took the evolution of the underlying disk structure including a likely range of disk lifetimes into account. We motivated our simulations from the available observational constraints of the planet and disk parameters. 

We find that the final masses and semi-major axes of the simulated population are largely determined by the $\alpha$ viscosity parameter of the underlying disk structure, for which we adopted three different values ($\alpha \in \{\num{1e-4}, \num{3.16e-4}, \num{1e-3}\}$). A higher $\alpha$ parameter results in higher migration and accretion rates of the embedded planets, leading to a more massive and close-in population with a clustering of planets near the inner edge of the disk for $\alpha = 10^{-3}$. In contrast, a lower viscosity parameter, $\alpha = 10^{-4}$, results in a less massive final population where the simulated planets remain mostly outside the inner region of $a \lesssim 10\,$AU and generally have masses below $3\,M_{\mathrm{Jup}}$. An intermediate value of $\alpha \sim 3\times 10^{-4}$ results in a blend of the two outcomes in terms of semi-major axis, while reaching intermediate masses. Only to a lesser extent do the final parameters correlate with the gas life-time of the disk or the initial positions of the planets.

If young planets with wide orbits, such as the putative planets of HD 163296, are the norm rather than an exception, we might expect them to eventually form the population of massive exoplanets that has been observed over the last decades. Under this assumption, we studied the synthesized giant planet population from our simulations in the context of known exoplanets of stars similar in size and luminosity to HD 163296. 

From the mass and semi-major axis distribution of the simulated planets we conclude that a high $\alpha$ viscosity parameter on the order of $\num{3.16e-4} \lesssim \alpha \lesssim 10^{-3}$ is most likely to produce a planet population that is consistent with current observations. Simulations with low viscosity, that is $\alpha = 10^{-4}$, did not generate a sufficient number of close-in planets to explain the observed abundance of giant planets on short-period orbits. We argue that future direct imaging observations with increased measurement sensitivity could further constrain the value of $\alpha$, still under the assumption that the employed initial population of wide orbit giants is typical of exoplanet systems. If a larger number of planets with $m_\mathrm{pl} \lesssim 3\, M_{\mathrm{Jup}}$ and $a_\mathrm{pl} \gtrsim 10\,$AU will be uncovered in future observations, their existence would point toward lower migration and accretion rates and an intermediate viscosity parameter, that is $\alpha < 10^{-3}$, whereas the lack thereof would imply even higher viscosities on the order of $\alpha \approx 10^{-3}$.

Our simulations show an increase of the orbital eccentricities with increasing $\alpha$ but even for the simulations with the highest $\alpha$ values the mean eccentricities of the simulated planets are still below those of the known giant planets. We attribute the moderate eccentricities to the comparably low number of planets and point out the growing evidence for a further planet in the system which would likely change the distribution. Although we do not expect to fully reproduce the eccentricity distribution of the observed exoplanets, since we are only modeling an example system, it is nevertheless intriguing that the eccentricities of the simulated planets are comparatively low. In addition, further studies are needed to determine whether most observed RV and direct imaging planetary systems form from the rings and gaps we observe in protoplanetary disks such as HD 163296.

In accordance with the results by \citet{Wagner2019} and \citet{Bitsch20}, we observe a discrepancy between the hypothetically observable and the actual multiplicities of our simulated systems. This implies that many stars known to have one giant planet might in fact harbor several more that are inaccessible to us at current measurement sensitivities.

(Near) Resonant motion is common among all our simulations and oftentimes acts as a stabilizing factor in closely spaced systems. This result is interesting for two reasons. Firstly it is in contrast to findings by \citet{Bitsch20} who found no evidence for resonant motion of giant planets. This seems to imply that the occurrence of resonances depends on the formation pathways, including the initial number and spacing of the accreting planetary embryos. Secondly, we suggest that the occurrence rates of resonances among detected exoplanets could be used to constrain the $\alpha$ viscosity parameter of the disks they originated from, under the assumptions that we made for the formation pathway. We base this on the fact that the number of observable resonant pairs in our simulations increases for higher values of the viscosity parameter $\alpha$.

Additional simulations and observations, especially of disks where several planets are expected, could provide valuable insights to constrain the different planet formation pathways even further. In particular, many future observational campaigns will be aimed at fully characterizing targeted systems of particular interest, rather than performing broad surveys of the sky. In this context, theoretical investigations of individual systems, such as this work on the putative planets around HD 163296, rather than of general representatives of the overall sample, represent an essential tool for making sense of the anticipated observational data and expanding our knowledge of planetary formation.

\begin{acknowledgements}

B.B. and G.P. thank the European Research Council (ERC Starting Grant 757448-PAMDORA) for their financial support. We also thank the MPIA summer internship program who helped to finance J.M.-H. stay at the MPIA.\\
This research has made use of the NASA Exoplanet Archive, which is operated by the California Institute of Technology, under contract with the National Aeronautics and Space Administration under the Exoplanet Exploration Program.

\end{acknowledgements}

% WARNING
%-------------------------------------------------------------------
% Please note that we have included the references to the file aa.dem in
% order to compile it, but we ask you to:
%
% - use BibTeX with the regular commands:
   \bibliographystyle{aa} % style aa.bst
   \bibliography{references.bib} % your references Yourfile.bib

\begin{thebibliography}{108}
\expandafter\ifx\csname natexlab\endcsname\relax\def\natexlab#1{#1}\fi

\bibitem[{{Alexander} {et~al.}(2014){Alexander}, {Pascucci}, {Andrews},
  {Armitage}, \& {Cieza}}]{Alexander}
{Alexander}, R., {Pascucci}, I., {Andrews}, S., {Armitage}, P., \& {Cieza}, L.
  2014, in Protostars and Planets VI, ed. H.~{Beuther}, R.~S. {Klessen}, C.~P.
  {Dullemond}, \& T.~{Henning}, 475

\bibitem[{{ALMA Partnership} {et~al.}(2015){ALMA Partnership}, {Brogan},
  {P{\'e}rez}, {Hunter}, {Dent}, {Hales}, {Hills}, {Corder}, {Fomalont},
  {Vlahakis}, {Asaki}, {Barkats}, {Hirota}, {Hodge}, {Impellizzeri}, {Kneissl},
  {Liuzzo}, {Lucas}, {Marcelino}, {Matsushita}, {Nakanishi}, {Phillips},
  {Richards}, {Toledo}, {Aladro}, {Broguiere}, {Cortes}, {Cortes}, {Espada},
  {Galarza}, {Garcia-Appadoo}, {Guzman-Ramirez}, {Humphreys}, {Jung}, {Kameno},
  {Laing}, {Leon}, {Marconi}, {Mignano}, {Nikolic}, {Nyman}, {Radiszcz},
  {Remijan}, {Rod{\'o}n}, {Sawada}, {Takahashi}, {Tilanus}, {Vila Vilaro},
  {Watson}, {Wiklind}, {Akiyama}, {Chapillon}, {de Gregorio-Monsalvo}, {Di
  Francesco}, {Gueth}, {Kawamura}, {Lee}, {Nguyen Luong}, {Mangum}, {Pietu},
  {Sanhueza}, {Saigo}, {Takakuwa}, {Ubach}, {van Kempen}, {Wootten},
  {Castro-Carrizo}, {Francke}, {Gallardo}, {Garcia}, {Gonzalez}, {Hill},
  {Kaminski}, {Kurono}, {Liu}, {Lopez}, {Morales}, {Plarre}, {Schieven},
  {Testi}, {Videla}, {Villard}, {Andreani}, {Hibbard}, \& {Tatematsu}}]{ALMA}
{ALMA Partnership}, {Brogan}, C.~L., {P{\'e}rez}, L.~M., {et~al.} 2015, \apjl,
  808, L3

\bibitem[{{Andrews} {et~al.}(2016){Andrews}, {Wilner}, {Zhu}, {Birnstiel},
  {Carpenter}, {P{\'e}rez}, {Bai}, {{\"O}berg}, {Hughes}, {Isella}, \&
  {Ricci}}]{Andrews}
{Andrews}, S.~M., {Wilner}, D.~J., {Zhu}, Z., {et~al.} 2016, \apjl, 820, L40

\bibitem[{{Anglada-Escud{\'e}} {et~al.}(2010){Anglada-Escud{\'e}},
  {L{\'o}pez-Morales}, \& {Chambers}}]{Anglada}
{Anglada-Escud{\'e}}, G., {L{\'o}pez-Morales}, M., \& {Chambers}, J.~E. 2010,
  \apj, 709, 168

\bibitem[{{Ataiee} {et~al.}(2018){Ataiee}, {Baruteau}, {Alibert}, \&
  {Benz}}]{Ataiee}
{Ataiee}, S., {Baruteau}, C., {Alibert}, Y., \& {Benz}, W. 2018, \aap, 615,
  A110

\bibitem[{{Ayliffe} \& {Bate}(2009)}]{Ayliffe}
{Ayliffe}, B.~A. \& {Bate}, M.~R. 2009, \mnras, 397, 657

\bibitem[{{Bach-M{\o}ller} \& {J{\o}rgensen}(2021)}]{Bach}
{Bach-M{\o}ller}, N. \& {J{\o}rgensen}, U.~G. 2021, \mnras, 500, 1313

\bibitem[{{Balbus} \& {Hawley}(1991)}]{Balbus1991}
{Balbus}, S.~A. \& {Hawley}, J.~F. 1991, \apj, 376, 214

\bibitem[{{Bitsch} {et~al.}(2013){Bitsch}, {Crida}, {Libert}, \&
  {Lega}}]{Bitsch13}
{Bitsch}, B., {Crida}, A., {Libert}, A.~S., \& {Lega}, E. 2013, \aap, 555, A124

\bibitem[{{Bitsch} {et~al.}(2019){Bitsch}, {Izidoro}, {Johansen}, {Raymond},
  {Morbidelli}, {Lambrechts}, \& {Jacobson}}]{Bitsch19}
{Bitsch}, B., {Izidoro}, A., {Johansen}, A., {et~al.} 2019, \aap, 623, A88

\bibitem[{{Bitsch} \& {Kley}(2011)}]{BitschKley}
{Bitsch}, B. \& {Kley}, W. 2011, \aap, 536, A77

\bibitem[{{Bitsch} {et~al.}(2018){Bitsch}, {Morbidelli}, {Johansen}, {Lega},
  {Lambrechts}, \& {Crida}}]{Bitsch18}
{Bitsch}, B., {Morbidelli}, A., {Johansen}, A., {et~al.} 2018, \aap, 612, A30

\bibitem[{{Bitsch} {et~al.}(2020){Bitsch}, {Trifonov}, \& {Izidoro}}]{Bitsch20}
{Bitsch}, B., {Trifonov}, T., \& {Izidoro}, A. 2020, \aap, 643, A66

\bibitem[{{Boisvert} {et~al.}(2018){Boisvert}, {Nelson}, \&
  {Steffen}}]{Boisvert2018}
{Boisvert}, J.~H., {Nelson}, B.~E., \& {Steffen}, J.~H. 2018, \mnras, 480, 2846

\bibitem[{{Boss}(2011)}]{Boss2011}
{Boss}, A.~P. 2011, \apj, 731, 74

\bibitem[{{Brauer} {et~al.}(2008){Brauer}, {Dullemond}, \& {Henning}}]{Brauer}
{Brauer}, F., {Dullemond}, C.~P., \& {Henning}, T. 2008, \aap, 480, 859

\bibitem[{{Calahan} {et~al.}(2021){Calahan}, {Bergin}, {Zhang}, {Schwarz},
  {Oberg}, {Guzman}, {Walsh}, {Aikawa}, {Alarcon}, {Andrews}, {Bae}, {Bergner},
  {Booth}, {Bosman}, {Cataldi}, {Czekala}, {Huang}, {Ilee}, {Law}, {Le Gal},
  {Long}, {Loomis}, {Menard}, {Nomura}, {Qi}, {Teague}, {van'T Hoff}, {Wilner},
  \& {Yamato}}]{Calahan}
{Calahan}, J.~K., {Bergin}, E.~A., {Zhang}, K., {et~al.} 2021, arXiv e-prints,
  arXiv:2109.06202

\bibitem[{{Chambers}(1999)}]{mercury}
{Chambers}, J.~E. 1999, \mnras, 304, 793

\bibitem[{{Chiang} \& {Goldreich}(1999)}]{Chiang}
{Chiang}, E.~I. \& {Goldreich}, P. 1999, \apj, 519, 279

\bibitem[{{Cimerman} {et~al.}(2017){Cimerman}, {Kuiper}, \&
  {Ormel}}]{Cimerman2017}
{Cimerman}, N.~P., {Kuiper}, R., \& {Ormel}, C.~W. 2017, \mnras, 471, 4662

\bibitem[{{Clarke} {et~al.}(2018){Clarke}, {Tazzari}, {Juhasz}, {Rosotti},
  {Booth}, {Facchini}, {Ilee}, {Johns-Krull}, {Kama}, {Meru}, \&
  {Prato}}]{Clarke}
{Clarke}, C.~J., {Tazzari}, M., {Juhasz}, A., {et~al.} 2018, \apjl, 866, L6

\bibitem[{{Coleman} \& {Nelson}(2016)}]{Coleman2016}
{Coleman}, G. A.~L. \& {Nelson}, R.~P. 2016, \mnras, 460, 2779

\bibitem[{{Cresswell} \& {Nelson}(2006)}]{Cresswell06}
{Cresswell}, P. \& {Nelson}, R.~P. 2006, \aap, 450, 833

\bibitem[{{Cresswell} \& {Nelson}(2008)}]{Cresswell08}
{Cresswell}, P. \& {Nelson}, R.~P. 2008, \aap, 482, 677

\bibitem[{{Crida} {et~al.}(2006){Crida}, {Morbidelli}, \& {Masset}}]{Crida}
{Crida}, A., {Morbidelli}, A., \& {Masset}, F. 2006, \icarus, 181, 587

\bibitem[{{D'Angelo} \& {Bodenheimer}(2013)}]{Angelo}
{D'Angelo}, G. \& {Bodenheimer}, P. 2013, \apj, 778, 77

\bibitem[{{Davies} {et~al.}(2014){Davies}, {Adams}, {Armitage}, {Chambers},
  {Ford}, {Morbidelli}, {Raymond}, \& {Veras}}]{Davies}
{Davies}, M.~B., {Adams}, F.~C., {Armitage}, P., {et~al.} 2014, in Protostars
  and Planets VI, ed. H.~{Beuther}, R.~S. {Klessen}, C.~P. {Dullemond}, \&
  T.~{Henning}, 787

\bibitem[{{Dipierro} {et~al.}(2015){Dipierro}, {Price}, {Laibe}, {Hirsh},
  {Cerioli}, \& {Lodato}}]{Dipierro}
{Dipierro}, G., {Price}, D., {Laibe}, G., {et~al.} 2015, \mnras, 453, L73

\bibitem[{{Duffell} \& {MacFadyen}(2013)}]{Duffell}
{Duffell}, P.~C. \& {MacFadyen}, A.~I. 2013, \apj, 769, 41

\bibitem[{{Dullemond} {et~al.}(2018){Dullemond}, {Birnstiel}, {Huang},
  {Kurtovic}, {Andrews}, {Guzm{\'a}n}, {P{\'e}rez}, {Isella}, {Zhu}, {Benisty},
  {Wilner}, {Bai}, {Carpenter}, {Zhang}, \& {Ricci}}]{Dullemond}
{Dullemond}, C.~P., {Birnstiel}, T., {Huang}, J., {et~al.} 2018, \apjl, 869,
  L46

\bibitem[{{Fedele} {et~al.}(2017){Fedele}, {Carney}, {Hogerheijde}, {Walsh},
  {Miotello}, {Klaassen}, {Bruderer}, {Henning}, \& {van Dishoeck}}]{Fedele}
{Fedele}, D., {Carney}, M., {Hogerheijde}, M.~R., {et~al.} 2017, \aap, 600, A72

\bibitem[{{Flaherty} {et~al.}(2017){Flaherty}, {Hughes}, {Rose}, {Simon}, {Qi},
  {Andrews}, {K{\'o}sp{\'a}l}, {Wilner}, {Chiang}, {Armitage}, \&
  {Bai}}]{Flaherty}
{Flaherty}, K.~M., {Hughes}, A.~M., {Rose}, S.~C., {et~al.} 2017, \apj, 843,
  150

\bibitem[{{Flock} {et~al.}(2015){Flock}, {Ruge}, {Dzyurkevich}, {Henning},
  {Klahr}, \& {Wolf}}]{Flock2015}
{Flock}, M., {Ruge}, J.~P., {Dzyurkevich}, N., {et~al.} 2015, \aap, 574, A68

\bibitem[{{Flock} {et~al.}(2019){Flock}, {Turner}, {Mulders}, {Hasegawa},
  {Nelson}, \& {Bitsch}}]{Flock}
{Flock}, M., {Turner}, N.~J., {Mulders}, G.~D., {et~al.} 2019, \aap, 630, A147

\bibitem[{{Fulton} {et~al.}(2021){Fulton}, {Rosenthal}, {Hirsch}, {Isaacson},
  {Howard}, {Dedrick}, {Sherstyuk}, {Blunt}, {Petigura}, {Knutson}, {Behmard},
  {Chontos}, {Crepp}, {Crossfield}, {Dalba}, {Fischer}, {Henry}, {Kane},
  {Kosiarek}, {Marcy}, {Rubenzahl}, {Weiss}, \& {Wright}}]{Fulton}
{Fulton}, B.~J., {Rosenthal}, L.~J., {Hirsch}, L.~A., {et~al.} 2021, \apjs,
  255, 14

\bibitem[{{Gaia Collaboration}(2018)}]{Gaia}
{Gaia Collaboration}. 2018, VizieR Online Data Catalog, I/345

\bibitem[{Gonzalez {et~al.}(2017)Gonzalez, Laibe, \& Maddison}]{Gonzalez}
Gonzalez, J.-F., Laibe, G., \& Maddison, S.~T. 2017, Monthly Notices of the
  Royal Astronomical Society, 467, 1984

\bibitem[{{Hara} {et~al.}(2019){Hara}, {Bou{\'e}}, {Laskar}, {Delisle}, \&
  {Unger}}]{Hara2019}
{Hara}, N.~C., {Bou{\'e}}, G., {Laskar}, J., {Delisle}, J.~B., \& {Unger}, N.
  2019, \mnras, 489, 738

\bibitem[{{Hartmann} {et~al.}(1998){Hartmann}, {Calvet}, {Gullbring}, \&
  {D'Alessio}}]{Hartmann}
{Hartmann}, L., {Calvet}, N., {Gullbring}, E., \& {D'Alessio}, P. 1998, \apj,
  495, 385

\bibitem[{{Huang} {et~al.}(2018){Huang}, {Andrews}, {Dullemond}, {Isella},
  {P{\'e}rez}, {Guzm{\'a}n}, {{\"O}berg}, {Zhu}, {Zhang}, {Bai}, {Benisty},
  {Birnstiel}, {Carpenter}, {Hughes}, {Ricci}, {Weaver}, \&
  {Wilner}}]{DSHARP-Huang}
{Huang}, J., {Andrews}, S.~M., {Dullemond}, C.~P., {et~al.} 2018, \apjl, 869,
  L42

\bibitem[{{H{\"u}hn} {et~al.}(2021){H{\"u}hn}, {Pichierri}, {Bitsch}, \&
  {Batygin}}]{Leon2021}
{H{\"u}hn}, L.~A., {Pichierri}, G., {Bitsch}, B., \& {Batygin}, K. 2021, \aap,
  656, A115

\bibitem[{{Isella} {et~al.}(2016){Isella}, {Guidi}, {Testi}, {Liu}, {Li}, {Li},
  {Weaver}, {Boehler}, {Carperter}, {De Gregorio-Monsalvo}, {Manara}, {Natta},
  {P{\'e}rez}, {Ricci}, {Sargent}, {Tazzari}, \& {Turner}}]{Isella+2016}
{Isella}, A., {Guidi}, G., {Testi}, L., {et~al.} 2016, \prl, 117, 251101

\bibitem[{{Isella} {et~al.}(2018){Isella}, {Huang}, {Andrews}, {Dullemond},
  {Birnstiel}, {Zhang}, {Zhu}, {Guzm{\'a}n}, {P{\'e}rez}, {Bai}, {Benisty},
  {Carpenter}, {Ricci}, \& {Wilner}}]{Isella+2018}
{Isella}, A., {Huang}, J., {Andrews}, S.~M., {et~al.} 2018, \apjl, 869, L49

\bibitem[{{Isella} {et~al.}(2007){Isella}, {Testi}, {Natta}, {Neri}, {Wilner},
  \& {Qi}}]{Isella+2017}
{Isella}, A., {Testi}, L., {Natta}, A., {et~al.} 2007, \aap, 469, 213

\bibitem[{{Izidoro} {et~al.}(2021){Izidoro}, {Bitsch}, {Raymond}, {Johansen},
  {Morbidelli}, {Lambrechts}, \& {Jacobson}}]{Izidoro21}
{Izidoro}, A., {Bitsch}, B., {Raymond}, S.~N., {et~al.} 2021, \aap, 650, A152

\bibitem[{{Izidoro} {et~al.}(2017){Izidoro}, {Ogihara}, {Raymond},
  {Morbidelli}, {Pierens}, {Bitsch}, {Cossou}, \& {Hersant}}]{Izidoro17}
{Izidoro}, A., {Ogihara}, M., {Raymond}, S.~N., {et~al.} 2017, \mnras, 470,
  1750

\bibitem[{{Johnson} {et~al.}(2010){Johnson}, {Aller}, {Howard}, \&
  {Crepp}}]{Johnson2010}
{Johnson}, J.~A., {Aller}, K.~M., {Howard}, A.~W., \& {Crepp}, J.~R. 2010,
  \pasp, 122, 905

\bibitem[{{Juri{\'c}} \& {Tremaine}(2008)}]{Juric2008}
{Juri{\'c}}, M. \& {Tremaine}, S. 2008, \apj, 686, 603

\bibitem[{{Kama} {et~al.}(2020){Kama}, {Trapman}, {Fedele, D.}, {Bruderer, S.},
  {Hogerheijde, M. R.}, {Miotello, A.}, {van Dishoeck, E. F.}, {Clarke, C.}, \&
  {Bergin, E. A.}}]{Kama}
{Kama}, M., {Trapman}, L., {Fedele, D.}, {et~al.} 2020, A\&A, 634, A88

\bibitem[{Kanagawa {et~al.}(2018)Kanagawa, Tanaka, \& Szuszkiewicz}]{Kanagawa}
Kanagawa, K., Tanaka, H., \& Szuszkiewicz, E. 2018, The Astrophysical Journal,
  861, 140

\bibitem[{{Kanagawa} {et~al.}(2015){Kanagawa}, {Tanaka}, {Muto}, {Tanigawa}, \&
  {Takeuchi}}]{Kanagawa15}
{Kanagawa}, K.~D., {Tanaka}, H., {Muto}, T., {Tanigawa}, T., \& {Takeuchi}, T.
  2015, \mnras, 448, 994

\bibitem[{{Keppler} {et~al.}(2018){Keppler}, {Benisty}, {M{\"u}ller},
  {Henning}, {van Boekel}, {Cantalloube}, {Ginski}, {van Holstein}, {Maire},
  {Pohl}, {Samland}, {Avenhaus}, {Baudino}, {Boccaletti}, {de Boer},
  {Bonnefoy}, {Chauvin}, {Desidera}, {Langlois}, {Lazzoni}, {Marleau},
  {Mordasini}, {Pawellek}, {Stolker}, {Vigan}, {Zurlo}, {Birnstiel},
  {Brandner}, {Feldt}, {Flock}, {Girard}, {Gratton}, {Hagelberg}, {Isella},
  {Janson}, {Juhasz}, {Kemmer}, {Kral}, {Lagrange}, {Launhardt}, {Matter},
  {M{\'e}nard}, {Milli}, {Molli{\`e}re}, {Olofsson}, {P{\'e}rez}, {Pinilla},
  {Pinte}, {Quanz}, {Schmidt}, {Udry}, {Wahhaj}, {Williams}, {Buenzli},
  {Cudel}, {Dominik}, {Galicher}, {Kasper}, {Lannier}, {Mesa}, {Mouillet},
  {Peretti}, {Perrot}, {Salter}, {Sissa}, {Wildi}, {Abe}, {Antichi},
  {Augereau}, {Baruffolo}, {Baudoz}, {Bazzon}, {Beuzit}, {Blanchard}, {Brems},
  {Buey}, {De Caprio}, {Carbillet}, {Carle}, {Cascone}, {Cheetham}, {Claudi},
  {Costille}, {Delboulb{\'e}}, {Dohlen}, {Fantinel}, {Feautrier}, {Fusco},
  {Giro}, {Gluck}, {Gry}, {Hubin}, {Hugot}, {Jaquet}, {Le Mignant}, {Llored},
  {Madec}, {Magnard}, {Martinez}, {Maurel}, {Meyer}, {M{\"o}ller-Nilsson},
  {Moulin}, {Mugnier}, {Orign{\'e}}, {Pavlov}, {Perret}, {Petit}, {Pragt},
  {Puget}, {Rabou}, {Ramos}, {Rigal}, {Rochat}, {Roelfsema}, {Rousset}, {Roux},
  {Salasnich}, {Sauvage}, {Sevin}, {Soenke}, {Stadler}, {Suarez}, {Turatto}, \&
  {Weber}}]{Keppler}
{Keppler}, M., {Benisty}, M., {M{\"u}ller}, A., {et~al.} 2018, \aap, 617, A44

\bibitem[{{Kley} \& {Dirksen}(2006)}]{Kley2006}
{Kley}, W. \& {Dirksen}, G. 2006, \aap, 447, 369

\bibitem[{{K{\"u}rster} {et~al.}(2015){K{\"u}rster}, {Trifonov}, {Reffert},
  {Kostogryz}, \& {Rodler}}]{Kuerster2015}
{K{\"u}rster}, M., {Trifonov}, T., {Reffert}, S., {Kostogryz}, N.~M., \&
  {Rodler}, F. 2015, \aap, 577, A103

\bibitem[{{Lambrechts} {et~al.}(2014){Lambrechts}, {Johansen}, \&
  {Morbidelli}}]{Lambrechts14}
{Lambrechts}, M., {Johansen}, A., \& {Morbidelli}, A. 2014, \aap, 572, A35

\bibitem[{{Lambrechts} {et~al.}(2019){Lambrechts}, {Lega}, {Nelson}, {Crida},
  \& {Morbidelli}}]{Lambrechts19}
{Lambrechts}, M., {Lega}, E., {Nelson}, R.~P., {Crida}, A., \& {Morbidelli}, A.
  2019, \aap, 630, A82

\bibitem[{{Lee} \& {Peale}(2002)}]{Lee}
{Lee}, M.~H. \& {Peale}, S.~J. 2002, \apj, 567, 596

\bibitem[{{Lega} {et~al.}(2015){Lega}, {Morbidelli}, {Bitsch}, {Crida}, \&
  {Szul{\'a}gyi}}]{Lega2015}
{Lega}, E., {Morbidelli}, A., {Bitsch}, B., {Crida}, A., \& {Szul{\'a}gyi}, J.
  2015, \mnras, 452, 1717

\bibitem[{{Lega} {et~al.}(2019){Lega}, {Morbidelli}, \& {Crida}}]{Lega2019}
{Lega}, E., {Morbidelli}, A., \& {Crida}, A. 2019, in EPSC-DPS Joint Meeting
  2019, Vol. 2019, EPSC--DPS2019--142

\bibitem[{{Li} {et~al.}(2021){Li}, {Chen}, {Lin}, \& {Zhang}}]{Li2021}
{Li}, Y.-P., {Chen}, Y.-X., {Lin}, D. N.~C., \& {Zhang}, X. 2021, \apj, 906, 52

\bibitem[{{Liu} {et~al.}(2018){Liu}, {Jin}, {Li}, {Isella}, \& {Li}}]{Liu}
{Liu}, S.-F., {Jin}, S., {Li}, S., {Isella}, A., \& {Li}, H. 2018, \apj, 857,
  87

\bibitem[{{Lodato} {et~al.}(2019){Lodato}, {Dipierro}, {Ragusa}, {Long},
  {Herczeg}, {Pascucci}, {Pinilla}, {Manara}, {Tazzari}, {Liu}, {Mulders},
  {Harsono}, {Boehler}, {M{\'e}nard}, {Johnstone}, {Salyk}, {van der Plas},
  {Cabrit}, {Edwards}, {Fischer}, {Hendler}, {Nisini}, {Rigliaco}, {Avenhaus},
  {Banzatti}, \& {Gully-Santiago}}]{Lodato}
{Lodato}, G., {Dipierro}, G., {Ragusa}, E., {et~al.} 2019, \mnras, 486, 453

\bibitem[{{Long} {et~al.}(2018){Long}, {Pinilla}, {Herczeg}, {Harsono},
  {Dipierro}, {Pascucci}, {Hendler}, {Tazzari}, {Ragusa}, {Salyk}, {Edwards},
  {Lodato}, {van de Plas}, {Johnstone}, {Liu}, {Boehler}, {Cabrit}, {Manara},
  {Menard}, {Mulders}, {Nisini}, {Fischer}, {Rigliaco}, {Banzatti}, {Avenhaus},
  \& {Gully-Santiago}}]{Long}
{Long}, F., {Pinilla}, P., {Herczeg}, G.~J., {et~al.} 2018, \apj, 869, 17

\bibitem[{{Lubow} \& {D'Angelo}(2006)}]{Lubow}
{Lubow}, S.~H. \& {D'Angelo}, G. 2006, \apj, 641, 526

\bibitem[{{Lynden-Bell} \& {Pringle}(1974)}]{Lynden}
{Lynden-Bell}, D. \& {Pringle}, J.~E. 1974, \mnras, 168, 603

\bibitem[{Machida {et~al.}(2010)Machida, Kokubo, Inutsuka, \&
  Matsumoto}]{Machida}
Machida, M.~N., Kokubo, E., Inutsuka, S.-i., \& Matsumoto, T. 2010, Monthly
  Notices of the Royal Astronomical Society, 405, 1227

\bibitem[{{Mamajek}(2009)}]{Mamajek}
{Mamajek}, E.~E. 2009, in American Institute of Physics Conference Series, Vol.
  1158, Exoplanets and Disks: Their Formation and Diversity, ed. T.~{Usuda},
  M.~{Tamura}, \& M.~{Ishii}, 3--10

\bibitem[{{Marois} {et~al.}(2008){Marois}, {Macintosh}, {Barman}, {Zuckerman},
  {Song}, {Patience}, {Lafreni{\`e}re}, \& {Doyon}}]{Marois2008}
{Marois}, C., {Macintosh}, B., {Barman}, T., {et~al.} 2008, Science, 322, 1348

\bibitem[{{Marois} {et~al.}(2010){Marois}, {Zuckerman}, {Konopacky},
  {Macintosh}, \& {Barman}}]{Marois2010}
{Marois}, C., {Zuckerman}, B., {Konopacky}, Q.~M., {Macintosh}, B., \&
  {Barman}, T. 2010, \nat, 468, 1080

\bibitem[{{Meru} {et~al.}(2019){Meru}, {Rosotti}, {Booth}, {Nazari}, \&
  {Clarke}}]{Meru}
{Meru}, F., {Rosotti}, G.~P., {Booth}, R.~A., {Nazari}, P., \& {Clarke}, C.~J.
  2019, \mnras, 482, 3678

\bibitem[{{Michel} {et~al.}(2021){Michel}, {van der Marel}, \&
  {Matthews}}]{Michel}
{Michel}, A., {van der Marel}, N., \& {Matthews}, B. 2021, arXiv e-prints,
  arXiv:2104.05894

\bibitem[{{Montesinos} {et~al.}(2009){Montesinos}, {Eiroa}, {Mora}, \&
  {Mer{\'\i}n}}]{Montesinos}
{Montesinos}, B., {Eiroa}, C., {Mora}, A., \& {Mer{\'\i}n}, B. 2009, \aap, 495,
  901

\bibitem[{{Morbidelli} \& {Nesvorny}(2012)}]{Morbidelli}
{Morbidelli}, A. \& {Nesvorny}, D. 2012, \aap, 546, A18

\bibitem[{{Nazari} {et~al.}(2019){Nazari}, {Booth}, {Clarke}, {Rosotti},
  {Tazzari}, {Juhasz}, \& {Meru}}]{Nazari}
{Nazari}, P., {Booth}, R.~A., {Clarke}, C.~J., {et~al.} 2019, \mnras, 485, 5914

\bibitem[{{Ndugu} {et~al.}(2019){Ndugu}, {Bitsch}, \& {Jurua}}]{Ndugu}
{Ndugu}, N., {Bitsch}, B., \& {Jurua}, E. 2019, \mnras, 488, 3625

\bibitem[{{Owen} {et~al.}(2013){Owen}, {Hudoba de Badyn}, {Clarke}, \&
  {Robins}}]{Owen}
{Owen}, J.~E., {Hudoba de Badyn}, M., {Clarke}, C.~J., \& {Robins}, L. 2013,
  \mnras, 436, 1430

\bibitem[{{Paardekooper} {et~al.}(2011){Paardekooper}, {Baruteau}, \&
  {Kley}}]{Paardekooper}
{Paardekooper}, S.~J., {Baruteau}, C., \& {Kley}, W. 2011, \mnras, 410, 293

\bibitem[{{Paardekooper} \& {Mellema}(2006)}]{Paardekooper2006}
{Paardekooper}, S.~J. \& {Mellema}, G. 2006, \aap, 453, 1129

\bibitem[{{Pinilla} {et~al.}(2012){Pinilla}, {Benisty}, \&
  {Birnstiel}}]{Pinilla}
{Pinilla}, P., {Benisty}, M., \& {Birnstiel}, T. 2012, \aap, 545, A81

\bibitem[{{Pinte} {et~al.}(2018){Pinte}, {Price}, {M{\'e}nard}, {Duch{\^e}ne},
  {Dent}, {Hill}, {de Gregorio-Monsalvo}, {Hales}, \& {Mentiplay}}]{Pinte}
{Pinte}, C., {Price}, D.~J., {M{\'e}nard}, F., {et~al.} 2018, \apjl, 860, L13

\bibitem[{{Rafikov}(2017)}]{Rafikov}
{Rafikov}, R.~R. 2017, \apj, 837, 163

\bibitem[{{Raymond} {et~al.}(2009){Raymond}, {Armitage}, \&
  {Gorelick}}]{Raymond2009}
{Raymond}, S.~N., {Armitage}, P.~J., \& {Gorelick}, N. 2009, \apjl, 699, L88

\bibitem[{{Raymond} {et~al.}(2010){Raymond}, {Armitage}, \&
  {Gorelick}}]{Raymond2009b}
{Raymond}, S.~N., {Armitage}, P.~J., \& {Gorelick}, N. 2010, \apj, 711, 772

\bibitem[{{Rodenkirch} {et~al.}(2021){Rodenkirch}, {Rometsch}, {Dullemond},
  {Weber}, \& {Kley}}]{Rodenkirch}
{Rodenkirch}, P.~J., {Rometsch}, T., {Dullemond}, C.~P., {Weber}, P., \&
  {Kley}, W. 2021, \aap, 647, A174

\bibitem[{{Ruge} {et~al.}(2016){Ruge}, {Flock}, {Wolf}, {Dzyurkevich},
  {Fromang}, {Henning}, {Klahr}, \& {Meheut}}]{Ruge}
{Ruge}, J.~P., {Flock}, M., {Wolf}, S., {et~al.} 2016, \aap, 590, A17

\bibitem[{{Schlaufman}(2018)}]{Schlaufman}
{Schlaufman}, K.~C. 2018, \apj, 853, 37

\bibitem[{{Schulik} {et~al.}(2019){Schulik}, {Johansen}, {Bitsch}, \&
  {Lega}}]{Schulik}
{Schulik}, M., {Johansen}, A., {Bitsch}, B., \& {Lega}, E. 2019, \aap, 632,
  A118

\bibitem[{{Shakura} \& {Sunyaev}(1973)}]{Shakura}
{Shakura}, N.~I. \& {Sunyaev}, R.~A. 1973, \aap, 500, 33

\bibitem[{{Sotiriadis} {et~al.}(2017){Sotiriadis}, {Libert}, {Bitsch}, \&
  {Crida}}]{Sotiriadis2017}
{Sotiriadis}, S., {Libert}, A.-S., {Bitsch}, B., \& {Crida}, A. 2017, \aap,
  598, A70

\bibitem[{Tanaka \& Ward(2004)}]{Tanaka2004}
Tanaka, H. \& Ward, W.~R. 2004, The Astrophysical Journal, 602, 388

\bibitem[{{Tanaka} {et~al.}(2022){Tanaka}, {Kanagawa}, {Tanaka}, \&
  {Tanigawa}}]{Tanaka2022}
{Tanaka}, Y.~A., {Kanagawa}, K.~D., {Tanaka}, H., \& {Tanigawa}, T. 2022, \apj,
  925, 95

\bibitem[{{Tanigawa} \& {Tanaka}(2016)}]{Tanigawa2016}
{Tanigawa}, T. \& {Tanaka}, H. 2016, \apj, 823, 48

\bibitem[{{Teague} {et~al.}(2021){Teague}, {Bae}, {Aikawa}, {Andrews},
  {Bergin}, {Bergner}, {Boehler}, {Booth}, {Bosman}, {Cataldi}, {Czekala},
  {Guzm{\'a}n}, {Huang}, {Ilee}, {Law}, {Le Gal}, {Long}, {Loomis},
  {M{\'e}nard}, {{\"O}berg}, {P{\'e}rez}, {Schwarz}, {Sierra}, {Walsh},
  {Wilner}, {Yamato}, \& {Zhang}}]{Teague2021}
{Teague}, R., {Bae}, J., {Aikawa}, Y., {et~al.} 2021, \apjs, 257, 18

\bibitem[{{Teague} {et~al.}(2019){Teague}, {Bae}, \& {Bergin}}]{Teague19}
{Teague}, R., {Bae}, J., \& {Bergin}, E.~A. 2019, \nat, 574, 378

\bibitem[{Teague {et~al.}(2018)Teague, Bae, Bergin, Birnstiel, \&
  Foreman-Mackey}]{Teague2018}
Teague, R., Bae, J., Bergin, E.~A., Birnstiel, T., \& Foreman-Mackey, D. 2018,
  860, L12

\bibitem[{{van der Marel} {et~al.}(2019){van der Marel}, {Dong}, {di
  Francesco}, {Williams}, \& {Tobin}}]{Marel}
{van der Marel}, N., {Dong}, R., {di Francesco}, J., {Williams}, J.~P., \&
  {Tobin}, J. 2019, \apj, 872, 112

\bibitem[{{Wagner} {et~al.}(2019){Wagner}, {Apai}, \& {Kratter}}]{Wagner2019}
{Wagner}, K., {Apai}, D., \& {Kratter}, K.~M. 2019, \apj, 877, 46

\bibitem[{{Walsh} {et~al.}(2014){Walsh}, {Juh{\'a}sz}, {Pinilla}, {Harsono},
  {Mathews}, {Dent}, {Hogerheijde}, {Birnstiel}, {Meeus}, {Nomura}, {Aikawa},
  {Millar}, \& {Sandell}}]{Walsh}
{Walsh}, C., {Juh{\'a}sz}, A., {Pinilla}, P., {et~al.} 2014, \apjl, 791, L6

\bibitem[{{Wang} {et~al.}(2018){Wang}, {Graham}, {Dawson}, {Fabrycky}, {De
  Rosa}, {Pueyo}, {Konopacky}, {Macintosh}, {Marois}, {Chiang}, {Ammons},
  {Arriaga}, {Bailey}, {Barman}, {Bulger}, {Chilcote}, {Cotten}, {Doyon},
  {Duch{\^e}ne}, {Esposito}, {Fitzgerald}, {Follette}, {Gerard}, {Goodsell},
  {Greenbaum}, {Hibon}, {Hung}, {Ingraham}, {Kalas}, {Larkin}, {Maire},
  {Marchis}, {Marley}, {Metchev}, {Millar-Blanchaer}, {Nielsen}, {Oppenheimer},
  {Palmer}, {Patience}, {Perrin}, {Poyneer}, {Rajan}, {Rameau},
  {Rantakyr{\"o}}, {Ruffio}, {Savransky}, {Schneider}, {Sivaramakrishnan},
  {Song}, {Soummer}, {Thomas}, {Wallace}, {Ward-Duong}, {Wiktorowicz}, \&
  {Wolff}}]{Wang2018}
{Wang}, J.~J., {Graham}, J.~R., {Dawson}, R., {et~al.} 2018, \aj, 156, 192

\bibitem[{{Weber} {et~al.}(2019){Weber}, {P{\'e}rez}, {Ben{\'\i}tez-Llambay},
  {Gressel}, {Casassus}, \& {Krapp}}]{Weber2019}
{Weber}, P., {P{\'e}rez}, S., {Ben{\'\i}tez-Llambay}, P., {et~al.} 2019, \apj,
  884, 178

\bibitem[{{Weidenschilling} \& {Marzari}(1996)}]{Weidenschilling}
{Weidenschilling}, S.~J. \& {Marzari}, F. 1996, \nat, 384, 619

\bibitem[{{Winn} \& {Fabrycky}(2015)}]{Winn}
{Winn}, J.~N. \& {Fabrycky}, D.~C. 2015, \araa, 53, 409

\bibitem[{{Wittenmyer} {et~al.}(2019){Wittenmyer}, {Clark}, {Zhao}, {Horner},
  {Wang}, \& {Johns}}]{Wittenmyer2019}
{Wittenmyer}, R.~A., {Clark}, J.~T., {Zhao}, J., {et~al.} 2019, \mnras, 484,
  5859

\bibitem[{{Wittenmyer} {et~al.}(2020){Wittenmyer}, {Wang}, {Horner}, {Butler},
  {Tinney}, {Carter}, {Wright}, {Jones}, {Bailey}, {O'Toole}, \&
  {Johns}}]{Wittenmyer2020}
{Wittenmyer}, R.~A., {Wang}, S., {Horner}, J., {et~al.} 2020, \mnras, 492, 377

\bibitem[{{Wittenmyer} {et~al.}(2013){Wittenmyer}, {Wang}, {Horner}, {Tinney},
  {Butler}, {Jones}, {O'Toole}, {Bailey}, {Carter}, {Salter}, {Wright}, \&
  {Zhou}}]{Wittenmyer2013}
{Wittenmyer}, R.~A., {Wang}, S., {Horner}, J., {et~al.} 2013, \apjs, 208, 2

\bibitem[{{Wright} {et~al.}(2011){Wright}, {Veras}, {Ford}, {Johnson}, {Marcy},
  {Howard}, {Isaacson}, {Fischer}, {Spronck}, {Anderson}, \&
  {Valenti}}]{Wright}
{Wright}, J.~T., {Veras}, D., {Ford}, E.~B., {et~al.} 2011, \apj, 730, 93

\bibitem[{{Zhang} {et~al.}(2015){Zhang}, {Blake}, \& {Bergin}}]{Zhang+2015}
{Zhang}, K., {Blake}, G.~A., \& {Bergin}, E.~A. 2015, \apjl, 806, L7

\bibitem[{{Zhang} {et~al.}(2018){Zhang}, {Zhu}, {Huang}, {Guzm{\'a}n},
  {Andrews}, {Birnstiel}, {Dullemond}, {Carpenter}, {Isella}, {P{\'e}rez},
  {Benisty}, {Wilner}, {Baruteau}, {Bai}, \& {Ricci}}]{Zhang2018}
{Zhang}, S., {Zhu}, Z., {Huang}, J., {et~al.} 2018, \apjl, 869, L47

\end{thebibliography}
%
% - join the .bib files when you upload your source files
%-------------------------------------------------------------------

\end{document}